 \documentclass[final,3p,times]{elsarticle}

\usepackage{amssymb}
\usepackage{amsmath}
\usepackage{graphicx}
\usepackage{multirow}
\usepackage{mathptmx}
\usepackage{isotope}

\journal{ABC}

\begin{document}

\begin{frontmatter}



\title{Influence of baryon number density-dependent bag function $B(n)$ on strange stars for non-zero value of strange quark mass ($m_s\neq0$) in $f(R,T)$ gravity consistent with observational results} 


\author[label1]{Rohit Roy} 
\author[label2]{Debadri Bhattacharjee}
\author[label3]{Koushik Ballav Goswami}
\author[label4]{Pradip Kumar Chattopadhyay}
\affiliation[label1,label2,label3,label4]{organization={IUCAA, Centre for Astronomy and Astrophysics (ICARD), Department of Physics, Coochbehar Panchanan Barma University},
            addressline={Panchanan Nagar, Vivekananda Street}, 
            city={Coochbehar},
            postcode={736101}, 
            state={West Bengal},
            country={India}}

\begin{abstract}
In this work, a new class of solution of the Einstein field equation for an isotropic strange star using the modified Mak-Harko type density profile along with the equation of state as proposed in the MIT bag model and considering finite mass of the strange quark ($m_s$) is presented in the framework of $f(R,T)$ gravity with $f(R,T)=R+2\zeta T$, where, $\zeta$ is the coupling parameter. To incorporate the quark matter hypothesis with a physically viable stellar framework, a baryon number density ($n$) dependent bag function $B(n)$ is analysed, using exponential type parametrisation. The energy per baryon ($E_B$) has been investigated to restrict $B(n)$ and corresponding $n$ within a stable window, specifically satisfying the condition $E_B\leq 930.4~MeV$, which corresponds to the binding energy of $\isotope[56]{Fe}$. We note a lower limit of $n$ below which $E_B>930.4~MeV$ as $E_B$ increases with the decrease of $n$. This value, however, depends on $m_s$. Additionally, $n$ has a maximum value of $0.36~fm^{-3}$ irrespective of $m_s$ depending on the range of bag function. All the essential characteristics are satisfactorily fulfilled within the stellar interior for the selected set of parameter space. In this model, the maximum mass and radius are found by solving the TOV equations numerically which yields $M=2.03~M_{\odot}$ with a radius of $11.49~km$ for $m_s=0~MeV$ and $n=0.36~fm^{-3}$ and $\zeta=-0.1$. It is also noted  that the maximum mass and the corresponding radius are the function of $m_s$, $\zeta$ and $n$. The proposed model has been shown to comply with the required energy conditions and satisfies the criterion for dynamical stability, thereby confirming its physical plausibility as a physically consistent stellar model within the parameter space used. 

\end{abstract}




\end{frontmatter}



\section{Introduction}\label{sec1}
General Relativity (GR) as proposed by Einstein has been a cornerstone in explaining gravitational phenomena on astrophysical and cosmological scales. It is in remarkable agreement with a wealth of weak gravity precision tests such as gravitational lensing of light from distant background stars, gravitational redshift, the Shapiro time-delay effect, anomalous perihelion precession of Mercury and Lunar laser experiments within the solar system. Outside the Solar system, predictions of GR involving detection of gravitational wave emission due to black hole (BH) mergers and changes in the orbit of binary pulsars are being verified \cite{IHS,BPA2,BPA3}. To examine the validity of GR in the strong gravity regime, one can analyse detections related to black hole mergers, which inherently involve extreme gravitational conditions. However, the question remains as to whether the GR holds good under such circumstances. The violation of the equivalence principle at black hole singularities raises concerns about the universality of GR, suggesting potential limitations to its applicability in these extreme environments. Therefore, despite its success, several unresolved issues in cosmology and astrophysics suggest that modifications to GR may be necessary. In this context Weyl \cite{HWE} first modified GR to unify gravitation and electromagnetism which is now followed by many researchers to modify GR to explain various astrophysical and cosmological aspects, such as the presence of dark energy, dark matter and late-time cosmic acceleration \cite{AGR,PER,BER,HAN,PJE,TPA}. In the regime of low-energy, both observational and theoretical challenges have emerged in the $\Lambda$CDM model, indicating potential limitations of GR as a fundamental description of gravity. These challenges necessitate the exploration of alternative gravitational frameworks \cite{LPE,JSB} that can remain consistent with observational data \cite{SNO,SMC,AAS,KBA,MRS,MJA,IHU}. A major issue with GR arises in the context of quantum field theory, where it lacks renormalizability within the framework of effective field theory \cite{ASH,JDO}. Specifically, GR cannot be quantised in a renormalizable manner without introducing an infinite number of independent parameters, which poses serious theoretical difficulties.\par

To address this non-renormalizability problem, modifications to the Einstein–Hilbert action have been proposed \cite{FBA}. One of the common approaches involve incorporating higher-order curvature invariants into the action, which become dominant at high energy scales and may resolve some shortcomings of GR. In addition to these modifications, various alternative gravitational models have been explored. Among them are scalar–tensor theories \cite{TSI}, which introduce additional scalar degrees of freedom to extend GR, teleparallel gravity \cite{SBA} which reformulates gravity using torsion instead of curvature, and other scalar field models \cite{GNO1,GNO2,CPA}. Furthermore, some theoretical frameworks aim to unify gravity with other fundamental interactions \cite{HWE,TKA,OKL}, whereas, others seek to incorporate quantum mechanical principles into the description of gravity \cite{RHO}. Given these theoretical inconsistencies and observational discrepancies, the search for a more complete gravitational theory beyond GR remains an active area of research, which is crucial for understanding the fundamental nature of spacetime and gravity.\par 

The $f(R)$ gravity model was introduced as an extension of GR to address several fundamental issues in cosmology and gravitational physics \cite{SCA}. In this framework, the standard Einstein–Hilbert action, which relies on the Ricci scalar $R$, is extended by replacing $R$ with a more general arbitrary function $f(R)$, which is a function of curvature scalar. This theory has been widely studied as a viable modification of GR to answer some puzzles related to accelerating universe and galactic dynamics. Harko et al. \cite{THA} extended the $f(R)$ gravity model by incorporating the trace of the energy-momentum tensor $T$, leading to the development of $f(R,T)$ gravity. In this framework, the function $f(R,T)$ depends not only on the Ricci scalar $R$ but also on $T$, which accounts for quantum effects and exotic imperfect fluids \cite{THA2}. A key outcome of their study was the demonstration that the covariant divergence of the energy-momentum tensor ($T_{\sigma\eta}$) does not vanish, resulting in an additional acceleration term owing to matter-curvature coupling. Consequently, in $f(R,T)$ gravity, test particles deviate from geodesic motion. However, Chakraborty \cite{SCH} later established that for a specific form, $f(R,T)=R+hT$, where $h$ is a constant, the geodesic motion of test particles is preserved. Numerous researchers \cite{MSH3,DMO} have explored the thermodynamic aspects of modified gravity. In the context of compact astrophysical objects, Moraes et al. \cite{PHR5} derived a hydrostatic equilibrium equation within the modified theory of $f(R,T)$ gravity. Their findings indicated that this  model allows for significantly higher maximum masses than the observed values. Several studies have demonstrated that $f(R,T)$ gravity can be applied to investigate gravitational waves \cite{MAL} and gravastars \cite{ADA2}. Rahaman et al. \cite{MRA} analysed the physical properties of compact objects in presence of anisotropy by incorporating the Karmakar condition within the $f(R,T)$ framework. Using Lie algebra and conformal Killing vectors, Das et al. \cite{ADA} constructed an analytical model for isotropic compact objects in $f(R,T)$ gravity, showing that the model accurately predicted the observed masses of various compact stars. A similar approach was adopted by Singh et al. \cite{KSH}, considering a suitable form of the function, $f(R,T)=R+2\zeta T$, where $\zeta$ could take both positive and negative values. Their results showed that smaller values of $\zeta$ lead to a stiffer equation of state (EoS). Carvalho et al. \cite{GAC3} investigated the mass-radius relationship of strange stars within this framework by considering the conservation of the energy-momentum tensor and analysing two distinct linear equation of state models. A number of authors \cite{PBH2,PBH3,PRE,MMO,DBH} have also modeled compact stars in different context of $f(R,T)$ gravity using different forms of metric potentials.\par

Compact astrophysical objects, such as Neutron Stars (NS) serve as ideal laboratories for testing the validity of gravitational theories beyond GR. These objects exhibit extreme densities and strong gravitational fields, providing an opportunity to explore deviations from Einstein gravity. In particular, NS harbour ultra-dense nuclear matter, whose equation of state (EoS) remains elusive. Any modification to gravitational theory could influence the mass-radius relationship, tidal deformability, and structural stability of these stars. In the context of $f(R,T)$ gravity, the inclusion of $T$ in the action induces additional contributions to the field equations, which may significantly impact the internal structure of compact objects. This modification introduces an effective energy-momentum tensor that alters the hydrostatic equilibrium conditions. The standard Tolman-Oppenheimer-Volkoff (TOV) equation, derived in GR, is consequently modified, leading to potential differences in observable parameters such as moment of inertia, tidal deformability, mass and radius.\par 

After the remarkable work of Chandrashekhar \cite{GSR,SCH2} the upper mass limit of a white dwarf star is known to be $1.4~M_{\odot}$. Das and Mukhopadhyay \cite{DDA} concluded that it may reach as high as $2.58$ $M_{\odot}$. However, in the case of NS, the maximum mass limit is still a debatable issue. Oppenheimer and Volkoff \cite{JRO} were among the first to calculate the mass limit of a NS, deriving an upper bound of $0.7~M_{\odot}$ under the assumption that neutrons are fermions. Rhoades and Ruffini \cite{CER} considered a perfect fluid with a density exceeding the nuclear density and, within the causal limit, estimated the maximum neutron star mass to be $3.2~M_{\odot}$. This limit has been extends to $3.6~M_{\odot}$ by Nauenberg and Chapline \cite{MNA}. Furthermore, Sabbadini and Hartle \cite{AGS} reported a maximum mass of $5~M_{\odot}$ for non-rotating NS with densities beyond the nuclear range, although their study did not impose causality constraints. Depending on the type of EoS, compact stars are proposed to have maximum masses within the range of $1.46-2.48~M_{\odot}$ \cite{PHA}. Typically the maximum mass of such objects with densities beyond the nuclear density is heavily EoS dependent, i.e., it depends on the internal composition of matter, which is yet to be accurately determined. In the last decade, the study of the internal composition of compact objects has attracted ample interest on the basis of different perspectives. One prominent hypothesis is the Quark Star theory, which suggests that NS may be composed of Quark Matter. This concept gained attention following recent observational findings \cite{EEM}. Itoh \cite{NIT} was the first to propose that quark stars could exist in a state of hydrostatic equilibrium. Madsen \cite{MAD} later demonstrated that quark stars composed solely of up ($u$) and down ($d$) quarks are inherently unstable if there is no external pressure. However, the inclusion of strange ($s$) quarks lowers the value of energy per baryon of the three flavour system compared to two-flavour ($u$, $d$) system, making the presence of $s$ quarks essential for stability. On the basis of such theory, a distinct class of compact stars, known as Strange Quark Stars (SQS) or strange star (SS) has been identified \cite{GBA,CAl,NKG}. Witten \cite{EWI} proposed that strange quark matter (SQM) might be the true ground state in the formalism of Quantum Chromodynamics (QCD) since its energy per baryon is lower than that of the most stable atomic nucleus, $\isotope[56]{\it Fe}$. This hypothesis further reinforces the idea that the SQM is a stable state. According to the asymptotic freedom property of the QCD, at extremely high densities, nucleons break down into their constituent quarks, forming a quark-gluon plasma (QGP) phase, where quarks interact weakly. Alford \cite{MAL2} suggested that the extreme density and low temperature in the core of NS could be sufficient requirement for the evolution of quark matter. However, the transition from hadronic matter to the phase of quark matter is not yet fully understood. Several phenomenological models attempt to describe this phase transition, and one widely used theoretical framework is the MIT bag model \cite{ACH}, which provides valuable insights into the properties of quark matter under extreme conditions. MIT bag model has been used in different formalism for investigating some important aspects of compact stars \cite{MS3,MS4,MS5,TN3,TN4}. \par

In the MIT bag model, the confinement of quarks is governed by the bag constant $B_g$, which represents the pressure difference between the perturbative and non-purterbative vacua. Now, in the SS modeling, $B_g$ has been considered as a constant parameter or to rely explicitly on density. For the constant $B_g$, the pressure difference between the two vacua remains constant, which may not be suitable for describing the different observed properties of compact objects, such as the mass-radius ratio, surface redshift, compactness etc. However, in the high density regime, it is imperative to consider the influence of the surrounding medium on which $B_g$ may impact the accurate representation of the EoS for the internal quark matter. In this context, it must be noted that as the density increases, the distinction between the perturbative and nonperturbative vacua may diminish, leading to the vanishing of $B_g$. Hence, considering a density dependent approach, we may obtain a flexible yet robust structure to model a wide range of SS candidates. Therefore, introducing a density dependent bag ($B_g(\rho)$) \cite{HRE} or a baryon number density dependent bag ($B_g(n)$) \cite{GFB} is more appropriate. Additionally, the inclusion of a nonzero strange quark mass ($m_s \neq 0$) is crucial in realistic SS modeling, as the presence of $m_s$ significantly modify the conventional SQS models while influencing their structural and dynamical properties. In the present model, we have considered the space-time to be static and spherically symmetric. Also, we have assumed that the stellar model is free from any charge and magnetic field. Under such assumptions the pressure isotropy condition may hold \cite{MS1,TN1,TN2,MS2}. The main aim of this work is to investigate properties of the strange stars that consist of $3$-flavour quarks and electrons in the formalism of $f(R,T)$ gravity for a specific density profile which is the modified form of density profile as prescribed by Mak and Harko \cite{MH}. Since both nonzero $m_s$ and baryon number density ($n$) have considerable effects on the energy per baryon ($E_B$) of the system, a nonzero $m_s$ along with a baryon number density dependent bag function ($B_g(n)$) is adopted in the study to construct a more realistic EoS to obtain the physical parameters of the stellar configuration. This approach facilitates the prediction of essential SS properties, including mass-radius relationships, stability conditions, causality constraints, and the validity of fundamental energy conditions.\par

The outline of this paper is as follows: In Sec.~\ref{sec2}, the thermodynamics between the constituents are analysed to obtain a relationship between density and pressure. Sec.~\ref{sec3} represents the baryon number density density dependent bag model by defining an appropriate functional dependence of the bag function ($B_g$) on the baryon number density $n$. The specific form of $B(n)$ considered in this study is analysed in detail. Furthermore, the relationship between the energy per baryon $E_B$ and the baryon number density $n$ is examined through graphical representations, which ultimately guide the selection of $B(n)$. The mathematical formulations of the $f(R,T)$ model are studied in Sec.~\ref{sec4}. In Sec.~\ref{sec5} modified Einstein field equations are obtained and solved for various parameters. Additionally, the proper bounds on the model parameters and the restriction on the coupling parameter $\zeta$ are addressed in this section. The maximum mass and radius relationship in the present model is studied in Sec.\ref{sec6} for various parametric combinations of model parameters. The physical viability of the proposed model is shown in Sec.~\ref{sec7} through radial variation of stellar parameters along with the study of causality and different energy conditions. The stability of the model is studied through various methods which are addressed in Sec.~\ref{sec8}. Finally, in Sec.~\ref{sec9}, the key findings of the model are discussed.

\section{Thermodynamical behaviour}\label{sec2}
In extremely dense and high-pressure environments, such as within compact objects, neutrons are thought to decompose into their constituent quarks. These quarks exist in equilibrium, ensuring charge neutrality of the stellar matter. The quarks collectively form a single colour-singlet baryon characterised by baryon number $A$. As quarks are fermionic in nature, quark matter can be described as a Fermi gas comprising $3A$ quarks. The effects of quark confinement can be effectively modeled within the framework of the bag model \cite{ACH,CHK}, governed by the following dynamical equations: 
\begin{equation}
	p=\sum_{i=u,d,s,e}p_i-B_g\label{1}
\end{equation}
\begin{equation}
	\rho=\sum_{i=u,d,s,e}\rho_i+B_g,\label{2}
\end{equation}
where $p_i$ and $\rho_i$ represent respectively the pressure and the energy density of the $i^{th}$ quark and electron and $B_g$, representing the effective difference between the perturbative and non perturbative vacuum, is termed as the bag constant. The presence of electrons along with up, down and strange quarks ensures the overall charge neutrality in the interior. Since presence of charm star is most likely unstable against small radial oscillations, the possibility of presence of the charm quark stars is forbidden \cite{CHK}. Muons are entirely absent in quark matter and only begin to appear at densities exceeding the critical density for the emergence of charm quarks. Therefore, the presence of muons can be neglected \cite{CHK}. Thus, in such a scenario the overall charge neutrality condition is given as:
\begin{equation}
	\sum_{i=u,d,s,e}  n_iq_i=0,\label{3}
\end{equation}
where, $q_i$ and $n_i$ are the charge and number density of the $i^{th}$ type particles, respectively. Now, the expressions for pressure ($p_i$), energy density ($\rho_i$) and number density ($n_i$) in the limit $T\rightarrow0$ are given as \cite{CHK,GXP}:
\begin{equation}
	p_i=\frac{g_i\mu^4\eta_i^4}{24\pi^2}\Big[\sqrt{1-\frac{m_i^2}{\mu_i^2}}\left(1-\frac{5m_i^2}{2\mu_i^2}\right)+\frac{3m_i^4}{2\mu_i^4}\log\frac{1+\sqrt{1-\frac{m_i^2}{\mu_i^2}}}{\frac{m_i}{\mu_i}}\Big],\label{4}
\end{equation}      
\begin{equation}
	\rho_i=\frac{g_i\mu^4\eta_i^4}{8\pi^2}\Big[\sqrt{1-\frac{m_i^2}{\mu_i^2}}\left(1-\frac{m_i^2}{2\mu_i^2}\right)-\frac{m_i^4}{2\mu_i^4}\log\frac{1+\sqrt{1-\frac{m_i^2}{\mu_i^2}}}{\frac{m_i}{\mu_i}} \Big],\label{5}
\end{equation}
\begin{equation}
	n_i=\frac{g_i\mu^3\eta_i^3}{6\pi^2}\Big[1-\frac{m_i^2}{\mu_i^2}\Big]^{3/2},\label{6}
\end{equation}
where, $g_i=2$ for leptons and $6$ for quarks. The charge neutrality condition is modified as:
\begin{equation}
	2\left(1-\frac{\mu_e}{\mu}\right)^3-\left(\frac{\mu_e}{\mu}\right)^3-\left(1-\frac{m_i^2}{\mu_i^2}\right)^{3/2}-1=0,\label{7}
\end{equation}
where, $m_i$ and $\mu_i$ represent the mass and the chemical potential of different species, respectively. In the high density core of compact objects, chemical equilibrium is established via the following weak interactions \cite{CHK},
\begin{equation}
	d\rightarrow u+e+\bar{\nu_{e}}, \;\;\;\;\;\;\;\;\; s\rightarrow u+e+\bar{\nu_{e}}, \label{8}
\end{equation}
\begin{equation}
	s+u \longleftrightarrow d+u. \label{9}
\end{equation}
In the formalism of a strange quark star, it is considered that such star may be composed of three flavour quarks ($u$, $d$ and $s$) and electrons. The mass of $u$, $d$ quarks as well as that of electrons are considerably lighter than the $s$ quarks. Therefore, $m_u=m_d=m_{e}=0$ can be set. Additionally, it follows from Eqs. (\ref{8}) and (\ref{9}), that $\mu_u=\mu-\mu_{e}$, where, $\mu_d=\mu_s=\mu$. Now, with the inclusion of non-zero strange quark mass ($m_s\neq0$), using  Eqs.~(\ref{4}), (\ref{5}) and (\ref{6}) a relationship between energy density ($\rho$) and pressure ($p$) of the system can be established in the following form:
\begin{equation}
	p=\frac{1}{3}(\rho-4B_1),\label{10}
\end{equation}
where, $B_{1}=\frac{4B_g+\rho_s-3p_s}{4}$, $p_s=$ pressure and $\rho_s$ is the energy density of the $s$ quark. It resembles the form of the basic MIT bag equation \cite{JKA} in the limit $m_s\rightarrow0$, for which $B_1\rightarrow{B_g}$.

\section{Baryon number density ($n$) dependent bag}\label{sec3}
The existence of massive NS with potential quark cores has been a topic of significant theoretical interest since the work of Itoh \cite{NIT} and continues to be an active area of investigation in current studies \cite{EAN}. While the EoS for the hadronic phase is relatively well-constrained, considerable uncertainty persists in describing the EoS for the quark phase. For accurate modeling, it is essential to formulate an EoS that characterise matter in situation where the core density significantly exceeds the nuclear saturation density. At such high densities, exotic components such as $\Delta$-isobars, hyperons, and mesonic contributions are anticipated to manifest, resulting a transition of phase from nuclear matter to a quark-gluon plasma (QGP) \cite{GBA,NKG,EWI}. A frequently employed framework for the quark phase is the MIT bag model EoS \cite{ACH,JKA}. The bag constant $B_g$, a key parameter in the MIT bag model, is determined to ensure compatibility with recent experimental findings from CERN regarding the formation of the QGP \cite{GFB}. Although experimental observations from CERN suggest that the QGP generated in heavy-ion collisions has a high temperature but negligible baryon number density. On the other hand the quark phase associated with strange star is characterized by a high baryon number density and low temperature. Initially, $B_g$ in the bag model was treated as a constant corresponding to its free-space value \cite{PAG}. Later, a phenomenological approach introduced a density-dependent $B_g$, incorporating its variation with the medium density \cite{DHL,XJI,XJI2,XJI3,HMU,HGU,PWA,LLZ}, which to some extent bridged the gap between theory and observational aspects. However, a rigorous QCD-based derivation of this density dependence for $B_g$ is still under investigation. The global colour symmetry model, which is an effective field theory of QCD, successfully explains the hadronic properties at temperature $T\rightarrow0$ and chemical potential $\mu\rightarrow0$ \cite{RTC,RTC2,CDR,MRF,PCT,XFL}. Various parameters related to the structure of hadrons such as, the bag constant have been introduced naturally. Liu et al. \cite{YXL} have investigated the density dependence of the bag constant of nucleons, the nucleon radius and the total energy of the bag as well as the quark condensate in nuclear matter in the global colour symmetry model, an effective field theory model of QCD. A maximal density of nuclear matter, which is larger than $12$ times the normal nucleon density, for the existence of the bag of quarks is obtained. As the maximal density is reached, a phase transition from nucleons to quark–gluons take place and the chiral symmetry is restored and the obtained changing features agree with those obtained in Quark Meson Coupling (QMC) and Relativistic Mean Field (RMF) quite well. In this sense, it provides a clue of the QCD foundation to the QMC and RMF with a simple effective field model of QCD. Liu et al. \cite{YXL} extended the global colour symmetry model proposed in Roberts et al. \cite{RTC2} to finite value of quark chemical potentials $\mu$, in order to derive the variation of $B$ with $\mu$ and $n$. Burgio et al. \cite{GFB} used the CERN SPS data of heavy-ion collisions to justify and determine the density dependence of B. Aguirre \cite{RA} used the NJL model to study the modification of the QCD vacuum with increasing baryonic density and to extract the relavent information about medium dependence of $B$. Liu et al. \cite{YXL} have not presented their $B(n)$ or $B(\mu)$ in a parametric form. Prasad and Bhalerao \cite{NPR} have fitted the results of Liu et al. \cite{YXL} with a suitable analytic expression with two parameters. We have used the expression of Prasad and Bhalerao \cite{NPR} in our work to study the properties of strange stars given below in the parametric form:  
\begin{equation}
	\frac{B(n)}{B_0}=e^{-(a_1x^2+a_2x)},\label{11}
\end{equation}
where, $x=\frac{n}{n_0}$ is the normalized baryon number density, $n_0$ is known as the baryon number density related to the ordinary nuclear matter ($n_0=0.17fm^{-3}$),~$a_1 = 0.0125657$,~$a_2=0.29522$ and $B_0=114~MeV/fm^3$. 
\begin{figure}[ht!]
	\centering
	\includegraphics{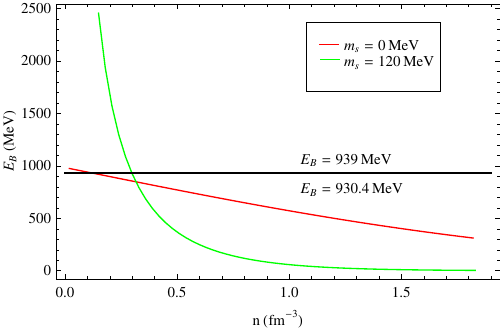}
	\caption{Variation of $E_B$ with $n$ for different parametric choices of $m_s$. The red and green lines represent the variations for $m_s=0$ and $120~MeV$, respectively.}\label{fig1}
\end{figure}

In this model, the value of $B_g$ is obtained form Eq.~(\ref{11}) as $B_g(n)=B(n)=B_{0}e^{-(a_1x^2+a_2x)}$ for a given baryon number density ($n$). Now, the energy per baryon ($E_B$) is crucial in determining the stability of a system composed of quarks. For the stability of three-flavour quark matter relative to $\isotope[56]{Fe}$, the energy per baryon ($E_B$) must be less than $930.4~MeV$, which represents the typical value of energy per baryon for $\isotope[56]{Fe}$. If $E_B$ falls within the range $930.4~MeV<E_B<939~MeV$, the system is considered to be composed of potentially metastable strange quark matter. However, for $E_B>939~MeV$, the quark matter becomes unstable \cite{MAD} i.e., it becomes confined to form stable hadronic matter. In Fig.~\ref{fig1}, the variation of $E_B$ with baryon number density ($n$) is plotted with a parametric choice of $m_s$. Fig.~\ref{fig1}, shows that, $E_B$ decreases with increasing $n$ for the parametric choice of $m_s$ and vice-versa. Several aspects can contribute to such a situation: $(i)$ baryons are held together by strong interaction, and with increasing $n$, the impact of strong interaction may dominate which may ultimately result in a lower $E_B$, $(ii)$ baryons being fermions; with increasing $n$, the Fermi phase space starts to fill up. As a result, the overall energy of the system might decrease due to interaction and energy balance of the system. Again, notably, from Fig.~\ref{fig1}, with increasing $n(n\geq0.3~fm^{-3})$ $E_B$ falls below $930.4~MeV$, which indicates that in such a high density regime, stable strange quark matter is bound to form. This further reinforces the quark star hypothesis. Thus for a given value of $m_s$, a lower bound of $n$ is obtained depending on various stability windows. In Table~\ref{tab1}, the bounds on $n$ are shown for different $m_s$ values. Notably, from Table~\ref{tab1}, the upper limit of $n$ is set owing to the fact that lower limit of bag should be greater than $57.55~MeV/fm^3$ irrespective of $m_s$. The value of $m_s$ lies normally within the range of $80~MeV<m_s<110~MeV$ as mentioned in the references \cite{MT,PDG}. However, for stability analysis of 3-flavour quark matter from energy per baryon, we have considered some lower as well as higher values of $m_s$. Such values beside this range are only weak constraint and within the acceptable QCD range.  
\begin{table}[ht!]
	\centering
	\caption{Range of baryon number density ($n$) in $fm^{-3}$ for different stability windows with a parametric choice of $m_s$ within the constraint for range of bag constant between $57.55~MeV/fm^3$ to $95.11~MeV/fm^3$.}\label{tab1}
	\resizebox{0.8\textwidth}{!}{$
		\begin{tabular}{@{}c|ccc}
			\hline
			$m_s$ &  stable window       &   metastable window        & unstable window \\
			$MeV$ &   $E_B<930.4~MeV$    &   $930.4~MeV<E_B<939~MeV$  & $E_B>939~MeV$  \\ \hline
			0     &    0.125$\leq$ n<~0.36 &   0.103$\leq$ n<~0.125        & n<~0.103       \\ 
			50    &    0.29$\leq$ n<~0.36  &   0.288 $\leq$ n<~0.29        & n<~0.288       \\
			100   &   0.296$\leq$ n<~0.36  &   0.295 $\leq$ n<~0.296       & n<~0.295       \\
			200   &    0.308$\leq$ n<~0.36  &   0.306 $\leq$ n<~0.308      & n<~0.306       \\ \hline
		\end{tabular}$}
\end{table}

\begin{figure}[ht!]
	\centering
	\includegraphics{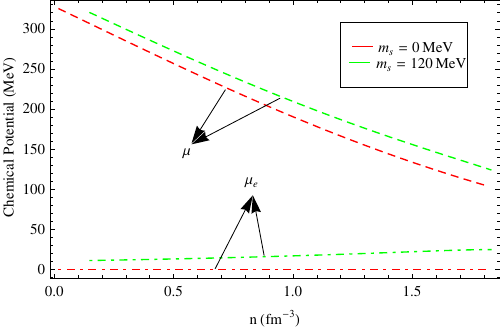}
	\caption{Variation of different chemical potentials with baryon number density. The red and green lines indicate the variation corresponding to $m_s=0$ and $120~MeV$, respectively.}\label{fig2}
\end{figure}
In Fig.~\ref{fig2}, the variation of different chemical potentials of quarks ($\mu$) and electrons ($\mu_e$) are shown. Notably, while $\mu$ decreases with increasing $n$ for $m_s\ge0$, $\mu_e$ on the other hand slightly increases for non-zero value of $m_s$. However, when $m_s=0$, the value of $\mu_e$ is also zero as evident from Fig.~\ref{fig2}. Although as $m_s\rightarrow 0$, $\mu_e\rightarrow 0$, to further solidify our choice of a density dependent bag, the variation of chemical potential of quarks and electrons with baryon number density ($n$) has been studied. In the case when $n$ is fixed, chemical potentials of quarks ($\mu$) and electrons ($\mu_e$) increase with the increase of $m_s$. The chemical potential is a key thermodynamic quantity that determines how the energy of a system changes when the number of particles changes. In the context of nuclear or high-energy physics, we are often concerned with the baryon chemical potential $\mu$, which tells us how much energy it costs to add one baryon to a system. In many physical systems, $\mu$ increases with increasing particle number density. In an ideal non-interacting Fermi gas (e.g., neutron gas): At zero temperature, the chemical potential equals the Fermi energy, which increases with density: $\mu=E_F\propto n^{2/3}$. So, $\mu$ increases with baryon number density. At finite temperature, the relationship is more complex, but still, in many standard systems $\mu$ increases with $n$.\par 
The statement that the chemical potential decreases with increasing baryon number density may seem counter-intuitive. So, to analyse this statement and give it a physical justification, we must consider the context and the equation of state (EoS) as well as thermodynamics of the system. There are certain physical conditions under which $\mu$ can decrease with increasing baryon number density. These include: 
\begin{enumerate}
	\item Strongly interacting systems with attractive interactions:
	\begin{itemize}
		\item If baryons attract each other (as in nuclear matter near saturation density), then adding more baryons can lower the average energy per particle. 
		\item This can cause the chemical potential to decrease with increasing $n$ in certain ranges.
		\item For example, in nuclear matter near the saturation point, the binding energy per nucleon becomes more negative as nucleons are added up to the saturation point. So, $\mu=(\frac{\partial E}{\partial n})$ decreases up to that point. 
	\end{itemize}
	\item First-order phase transitions:
	\begin{itemize}
		\item During a first-order phase transition (e.g., from hadronic matter to quark matter), the system can maintain constant chemical potential while the density increases (mixed phase).
		\item In some cases, small increases in density may lead to a drop in $\mu$ due to the change in phase composition. 
	\end{itemize}
	\item Thermodynamic Justification:
	\begin{itemize}
		\item From thermodynamics, the chemical potential is: $\mu=(\frac{\partial F}{\partial n})_{T,V}$, where, $F$ is the Gibbs free energy. If adding particles lowers the free energy (due to attractive interactions or entropy gain), then $\mu$ can decrease with increasing $n$. 
		\item From the Gibbs–Duhem relation: $d\mu=\frac{1}{n}(dp-sdT)$. At constant temperature, $(\frac{d\mu}{dn})$ can be negative if pressure increases slowly with density or decreases.
	\end{itemize}
\end{enumerate}
Moreover, heavier quarks require more energy to be accommodated. Hence, $\mu$ increases with increasing $m_s$ for a particular value of $n$. With increasing $n$, the lower Fermi levels are filled, and new electrons are pushed on to higher energy levels. As, $m_s$ increases the Fermi pressure increases. These two effects increase the energy cost of the system for the addition of new baryons; hence, $\mu_e$ increases with increasing $n$ and $m_s$. 

\section{Mathematical foundational field equations in $f(R,T)$ theory of gravity}\label{sec4}
The action in $f(R,T)$ formalism following the work of \cite{THA} is given as:
\begin{equation}
	S=\alpha\int d^4x\sqrt{-g}f(R,T)+\int d^4xL_m\sqrt{-g},\label{12}
\end{equation}
where, $\alpha=\frac{c^4}{16\pi G}$, $R$ is known as Ricci scalar, $T$ is the trace of the energy momentum tensor, $L_m$ is the Lagrangian density of matter and $g=\det {g_{\sigma\eta}}$. For the complete analysis of the model the condition that $8\pi G=1$ and $c=1$ is chosen for simplicity, which implies that $\alpha=\frac{1}{2}$. $L_m$ is intrinsically linked to the energy-momentum tensor $T_{\mu\nu}$ and is expressed as \cite{LDL}:
\begin{equation}
	T_{\sigma\eta}=-\frac{2}{\sqrt{-g}}\frac{\delta(\sqrt{-g}L_m)}{\delta g^{\sigma\eta}}.\label{13}
\end{equation}
If $L_m$ depends solely on the components of metric tensor $g_{\sigma\eta}$ and not on their derivatives, then Eq.~(\ref{13}) can be written as \cite{LDL}:
\begin{equation}
	T_{\sigma\eta}=L_mg_{\sigma\eta}-2\frac{\delta L_m}{\delta g^{\sigma\eta}}.\label{14}
\end{equation} 
Now, the variation of the action given in Eq.~(\ref{12}) with respect to the components of metric tensor produces the modified form of field equation in formalism of $f(R,T)$ gravity, which is given as:
\begin{equation}
	f_R(R,T)\Big[R_{\sigma\eta}+g_{\sigma\eta}\Box -\nabla_{\sigma}\nabla_{\eta}\Big]-\frac{1}{2}g_{\sigma\eta}f(R,T)=T_{\sigma\eta}-f_T(R,T)T_{\sigma\eta}-f_T(R,T)\Theta_{\sigma\eta},\label{15}
\end{equation}
where, $f_R(R,T)=\frac{\partial f(R,T)}{\partial R}$, $f_T(R,T)=\frac{\partial f(R,T)}{\partial T}$, $\Box \equiv \frac{1}{\sqrt{-g}}\partial_\mu(\sqrt{-g}g^{\sigma\eta}\partial_\nu)$, being the d'Alembertian operator, $\nabla_\sigma$ is described by the covariant derivative associated with the Levi-Civita connection of $g_{\sigma\eta}$ and $\Theta_{\sigma\eta}=g^{\alpha\beta}\frac{\delta T_{\alpha\beta}}{\delta g^{\sigma\eta}}$.
In this model of compact object, an isotropic environment with the following energy momentum tensor is considered,
\begin{equation}
	T_{\sigma\eta}=(\rho+p)\kappa_\sigma\kappa_\eta+p g_{\sigma\eta},\label{16}
\end{equation} 
where, $\rho$ and $p$ represent the energy density and the isotropic pressure, $\kappa^{\sigma}$ is the four velocity associated with the fluid satisfying $\kappa_{\sigma}\kappa^{\sigma}=-1$. Different studies have shown that two distinct choices are possible for the matter Lagrangian $(i)$ $L_m=p$, and $(ii)$ $L_m=-\rho$. According to the investigation of Faraoni \cite{VFA}, both the choices of Lagrangian are equivalent, for minimal matter-gravity coupling. In this model, according to the work of \cite{JMZ1} and \cite{FGA} the choice of $L_m=p$ is considered. The quantities $-\rho$ and $p$ are thermodynamically interchangeable. Harko et al. \cite{THA} demonstrated that representing the Lagrangian in terms of $-\rho$ or $p$ is entirely equivalent. In this work, adhering to the methodology outlined by Harko et al. \cite{THA}, the particular form of $f(R,T)=R+2\zeta T$ is considered, where $\zeta$ represents the coupling parameter. Constraints on $\zeta$ are necessary to ensure physically realistic models across astrophysical and cosmological domains. $\zeta$ can be limited to smaller values for the presence of a neutron star crust \cite{RLO}. While stellar equilibrium in white dwarfs imposes $\zeta > -3.0\times10^{-4}$ \cite{GAC}, considering the dark energy density parameter Ref.~\cite{SBH} proposed $\zeta \geq -1.9\times 10^{-8}$. Velten and Caram$\hat{e}$s \cite{HVE} further restricted $\zeta$ in the range $-0.1<\zeta<1.5$ on the basis of cosmological background evolution. The specific form of $f(R,T$) gravity has been widely and successfully applied in the literature \cite{HSA,HSA2,VSI,PHR,PHR2,PHR3,PHR4,MSH,DRK,PKU,MFS,KBG}. Many authors have used positive \cite{ADA,SSA,MSH2,AKY,PBH} and negative value \cite{DDE,GAC2} of coupling parameter $\zeta$. 
 
\section{Modified Einstein Field Equation and its solution in $f(R,T)$ gravity}\label{sec5}
Using a spherically symmetric and static metric,
\begin{equation}
ds^2=-e^{\nu(r)}dt^2+e^{\lambda(r)}dr^2+r^2(d{\theta^2}+\sin^2{\theta}d{\phi^2}),\label{17}
\end{equation}
where, $\nu(r)$ and $\lambda(r)$ are the unknown metric potentials and using Eqs.~(\ref{15}) and (\ref{16}) along with the choice of $f(R,T)=R+2\zeta T$, the following set of modified field equations are obtained:
\begin{equation}
	\frac{\lambda^{\prime}e^{-\lambda}}{r}+\frac{(1-e^{-\lambda})}{r^2}=(1+3\zeta)\rho-\zeta p,\label{18}
\end{equation}
\begin{equation}
	\frac{\nu^{\prime}e^{-\lambda}}{r}-\frac{(1-e^{-\lambda})}{r^2}=(1+3\zeta)p-\zeta\rho,\label{19}
\end{equation}
\begin{equation}
	e^{-\lambda}\left[\frac{\nu^{\prime\prime}}{2}+\frac{{\nu^{\prime}}^2}{4}+\frac{\nu^{\prime}-\lambda^{\prime}}{2r}-\frac{\nu^{\prime}\lambda^{\prime}}{4}\right]=(1+3\zeta)p-\zeta\rho.\label{20}
\end{equation}
where, ($\prime$) denotes differentiation with respect to $r$. To solve the modified field equations, we consider the form of density profile ($\rho$) as outlined below:
\begin{equation}
	\rho(r)=\rho_c\left[1-\left(1-\frac{\rho_0}{\rho_c}\right)\frac{r^2}{R^2}\right]+\zeta\rho_c\left(1-\frac{r^2}{R^2}\right),\label{21}
\end{equation} 
where $\rho_0$ and $\rho_c$ are the surface and central density, respectively. This density profile is regular, monotonically decreasing from centre to surface, satisfying all the physical plausibility criterion. Also when $\zeta=0$ it reduces to the original form as given by Mak and Harko \cite{MH}. Now, using Eqs.~(\ref{10}), (\ref{18}), (\ref{19}) and (\ref{21}), the solutions of the field equations in $f(R,T)$ formalism are given below:
\begin{eqnarray}
	\lambda(r)=-log\Big[\frac{1}{3rR^2}\Big(3rR^2-\frac{4B_1 r^3R^2\zeta}{3}-\frac{3r^5\rho_0}{5}-\frac{8\zeta r^5 \rho_0}{5}+\frac{3r^5\rho_c}{5}-r^3R^2\rho_c+\frac{8r^5\zeta^2\rho_c}{5}-\frac{8r^3R^2\zeta^2\rho_c}{3}\nonumber\\ +\frac{11}{5}r^5\zeta\rho_c-\frac{11}{3}r^3R^2\zeta\rho_c\Big)\Big],\label{22}
\end{eqnarray} 
\begin{eqnarray}
	\nu(r)=log\Bigg[1-\frac{R^{2}}{45}f_{3}\Bigg]+\frac{1}{(3+8\zeta)f_{2}}\Bigg[f_{1}\;tanh^{-1}\Bigg(\frac{R\{20B_1\zeta-(3+8\zeta)(\rho_{c}(1+\zeta)-6\rho_{0})\}}{\sqrt{5}f_{2}}\Bigg)\nonumber\\-f_{1}\;tanh^{-1}\Bigg(\frac{20B_1R^{2}\zeta+(3+8\zeta)(5R^{2}(1+\zeta)\rho_{c}-6r^{2}(\rho_{c}(1+\zeta)-\rho_{0}))}{\sqrt{5}Rf_{2}}\Bigg)+2f_{4}(1+\zeta)\Big[log(R^{4}f_{3}-45R^{2})\nonumber\\-log\{5R^{2}(r^{2}(4B_1\zeta+(1+\zeta)(3+8\zeta)\rho_c)-9)-3r^{4}(3+8\zeta)(\rho_{c}(1+\zeta)-\rho_{0})\}\Big] \Bigg],\label{23}
\end{eqnarray}
where, 
\begin{equation}
	f_1=2\sqrt{5}R(1+2\zeta)\Big(2B_1(9+34\zeta)-(1+\zeta)(3+8\zeta)\rho_c\Big),\label{24}
\end{equation}
\begin{eqnarray}
	\;\;\;\;\;\;\;\;\;\;\;f_2=\Bigg[80B_{1}^2R^2\zeta^2+40B_{1}R^2\zeta(1+\zeta)(3+8\zeta)\rho_c+(3+8\zeta)\Big(108\rho_0\nonumber\\+(1+\zeta)\rho_c\left(5R^2(1+\zeta)\rho_c(3+8\zeta)-108\right)\Big) \Bigg]^{\frac{1}{2}},\label{25}
\end{eqnarray} 
\begin{equation}
	\!\!\!\!\!\!\!\!\!\!\!f_3=20B_{1}\zeta+(3+8\zeta)(3\rho_0+2(1+\zeta)\rho_c),\label{26}
\end{equation}
and 
\begin{equation}
	f_4=\Bigg[80B_{1}^2R^2\zeta^2+108(3+8\zeta)\rho_0+4(1+\zeta)(3+8\zeta)\{10B_{1}R^2\zeta-27\}\rho_{c}+5R^2(1+\zeta)^2(3+8\zeta)^2\rho_{c}^2\Bigg]^{\frac{1}{2}}.\label{27}
\end{equation}
In Eq.~(\ref{22}) and (\ref{23}) dimension of $R$ is $km$, the dimensions of central density $(\rho_c)$ and surface density $(\rho_0)$ are both $km^{-2}$, the dimension of $B_1$ is $km^{-2}$ and $\zeta$ is dimensionless quantity. Accordingly the dimensions of $\lambda$ and $\nu$ become $(Length)^0$ as required and therefore mathematically consistent. 
\subsection{Bounds on different model parameters}\label{sec5.1}
According to the recent work of Goswami et al. \cite{KBG}, although the gravity is modified, the specific choice of $f(R,T)$ smoothly establishes that the extrinsic curvature tensor is zero, leading to the Darmois junction condition \cite{GDA} of GR. In GR, for any physically viable stellar model, the following conditions must be satisfied both within the stellar interior and at its boundary:
\begin{enumerate}
	\item {\it Boundary matching with the Schwarzschild exterior solution:}\\ At the stellar surface, the interior metric must smoothly match the Schwarzschild exterior solution, expressed as:
	\begin{equation}
		ds^2=-\left(1-\frac{2M}{r}\right)dt^2+\left(1-\frac{2M}{r}\right)^{-1}dr^2  +r^2\left(d{\theta^2}+\sin^2{\theta}d{\phi^2}\right),\label{28}
	\end{equation}
	where $M$ represents the total mass of the star.
	\item {\it Regularity of metric potentials and physical quantities:}\\ The metric potentials and the associated physical parameters characterising the stellar configuration must be finite, continuous, and well-behaved across the stellar interior, including at the centre.
	\item  {\it Boundary Condition on Radial Pressure:}\\ The total or effective pressure should decrease as the radial distance increases from the centre to the surface and must become zero at the surface, i.e., $p(r=R)=0$ \cite{PMA}.
	\item {\it Monotonicity and positivity of physical quantities:}\\ The value of energy density ($\rho$) and pressure ($p$) should be continuous and monotonically decreasing from the centre ($r=0$) to the surface ($r=R$). Furthermore, the ratio of pressure to energy density must comply with the Zeldovich condition, i.e., $\frac{p}{\rho}\leq1$ \cite {YBZ}.
	\item {\it Causality in sound propagation:}\\ The sound speed, represented by $v^2=\frac{dp}{d\rho}$ must obey the causality condition $0<v^2\leq 1$.
\end{enumerate}

\subsection{Restriction on the coupling parameter}\label{sec5.2}
The value of total active gravitational mass ($M$) contained within radius $R$ in $f(R,T)$ gravity can be obtained from boundary matching condition. Now, using Eqs.~(\ref{17}), (\ref{22}) and (\ref{28}) the total active gravitational mass is:
\begin{equation}
	M=\frac{R^3}{90}\Big[20B_{1}\zeta+(3+8\zeta)(3\rho_0+2(1+\zeta)\rho_c)\Big].\label{29}
\end{equation} 
For a physically viable stellar configuration the active gravitational mass ($M$) contained within radius $R$ should be always positive. Thus, from Eqs.~(\ref{29}), imposing the condition that ($i$) $\rho_c>0$ and ($ii$) $\rho_c>\rho_0$ the following allowed range for the coupling parameter ($\zeta$) is obtained:
\begin{equation}
	-\frac{3}{8}<\zeta<\frac{45M-18B_{1}R^3}{58B_{1}R^3}.\label{30}
\end{equation} 
and 
\begin{equation}
	-\frac{3}{8}<\zeta<\frac{\sqrt{3}}{32}\sqrt{\frac{480M+547B_{1}R^3}{B_{1}R^3}}-\frac{51}{32}.\label{31}
\end{equation}
Eqs.~(\ref{30}) and (\ref{31}), indicate that the range of $\zeta$ is highly mass and radius specific. Therefore, the chosen value of coupling ($\zeta$) should be within both the limit for a physically viable stellar structure. In Mak-Harko type density profile, the central and surface densities are variable quantity which must be solved using the proper boundary conditions. Now, the total mass which depends on energy density $(\rho)$ and radius $(R)$ should be positive. So, for such condition the central density $(\rho_c)$ can be obtained if the mass $(M)$ and radius (R) are given. Thus, for positive central density a limit on coupling may be obtained depending on mass and radius. Since the stellar model obtained in this article is geometry dependent, a universal bound on coupling parameter $(\zeta)$ may not be possible. In this model the theoretical constraint on lower limit of $\zeta$ is $-\frac{3}{8}(=-0.375)$. To address the physical features of the model with negative value of $\zeta$, we choose $\zeta=-0.1,~-0.2$ and $0.2$ as arbitrary choices to study the impact of coupling parameter $\zeta$ on gross properties of star. 

\section{Maximum mass from the TOV equation}\label{sec6}
\begin{figure}[ht!]
	\centering
	\begin{minipage}{0.48\textwidth}
		\centering
		\includegraphics[width=0.95\linewidth,height=0.75\textwidth]{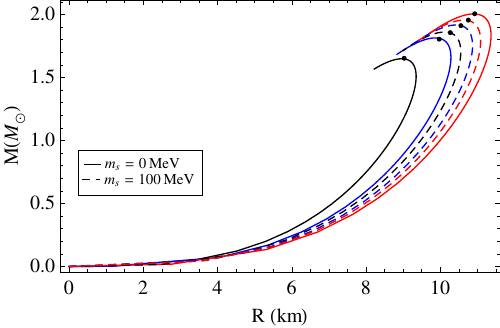}
		\caption{Mass-radius relation for a parametric choice of $\zeta=-0.1$. Here the solid black, blue and red lines represent the $M-R$ variation for $n=0.15~fm^{-3}$, $0.25~fm^{-3}$ and $0.35~fm^{-3}$, respectively with $m_s=0~MeV$. The dashed black, blue and red lines represent the corresponding variation for $n=0.3~fm^{-3}$, $0.33~fm^{-3}$ and $0.35~fm^{-3}$, respectively with $m_s=100~MeV$.}\label{fig3}
	\end{minipage}\hfil
	\begin{minipage}{0.48\textwidth}
		\centering
		\includegraphics[width=0.95\linewidth,height=0.75\textwidth]{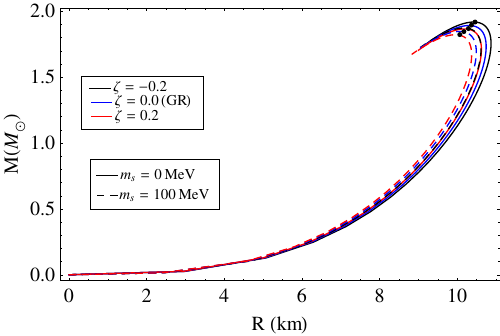}
		\caption{Mass-radius relation for a parametric choice of $n=0.3~fm^{-3}$. Here the solid black, blue and red lines represent the $M-R$ variation for  $\zeta=-0.2$, $0.0$ and $0.2$, respectively for $m_s=0~MeV$. The dashed black, blue and red lines represent the corresponding variation for $\zeta=-0.2$, $0.0$ and $0.2$, respectively with $m_s=100~MeV$.}\label{fig4}
	\end{minipage}\hfil
\end{figure}

\begin{figure}[ht!]
	\centering
	\begin{minipage}{0.45\textwidth}
		\centering
		\includegraphics[width=0.95\linewidth,height=0.65\textwidth]{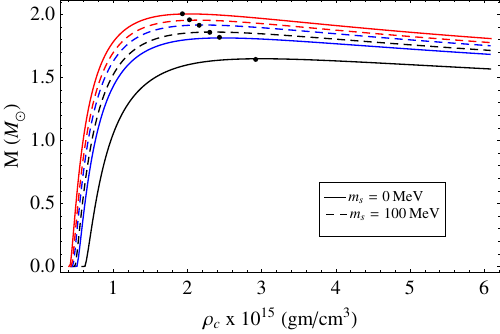}
		\caption{Variation of mass ($M$) with central density ($\rho_c$) for a parametric choice of $\zeta=-0.1$. Here the solid black, blue and red lines represent the $M-\rho_c$ variation for $n=0.15~fm^{-3}$, $0.25~fm^{-3}$ and $0.35~fm^{-3}$, respectively with $m_s=0~MeV$. The dashed black, blue and red lines represent the corresponding variation for $n=0.3~fm^{-3}$, $0.33~fm^{-3}$ and $0.35~fm^{-3}$, respectively with the choice $m_s=100~MeV$.}\label{fig5}
	\end{minipage}\hfil
	\begin{minipage}{0.45\textwidth}
		\centering
		\includegraphics[width=0.95\linewidth,height=0.65\textwidth]{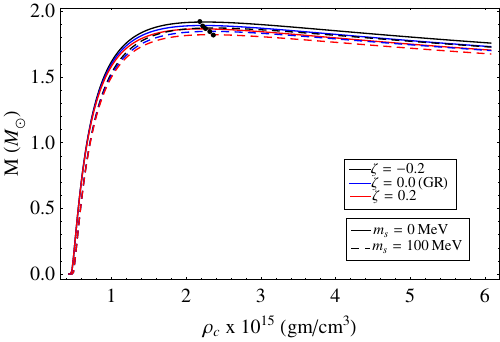}
		\caption{Variation of mass ($M$) with central density ($\rho_c$) for a parametric choice of $n=0.3~fm^{-3}$. Here the solid black, blue and red lines represent the $M-\rho_c$ variation for $\zeta=-0.2$, $0.0$ and $0.2$, respectively using $m_s=0~MeV$. The dashed black, blue and red lines represent the corresponding variation for $\zeta=-0.2$, $0.0$ and $0.2$, respectively, when $m_s=100~MeV$.}\label{fig6}
	\end{minipage}\hfil
\end{figure}

\begin{figure}[ht!]
	\centering
	\begin{minipage}{0.45\textwidth}
		\centering
		\includegraphics[width=0.95\linewidth,height=0.65\textwidth]{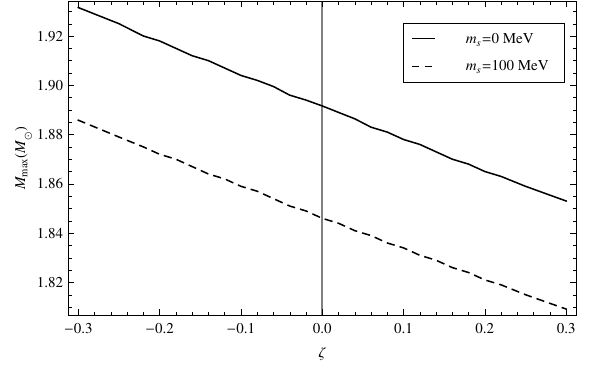}
		\caption{Variation of maximum mass ($M_{max}$) with coupling parameter $\zeta$ for a parametric choice of $n=0.3~fm^{-3}$. The solid and dashed lines represent the variation for $m_s=0~MeV$ and $m_s=100~MeV$, respectively. }\label{fig7}
	\end{minipage}\hfil
	\begin{minipage}{0.45\textwidth}
		\centering
		\includegraphics[width=0.95\linewidth,height=0.65\textwidth]{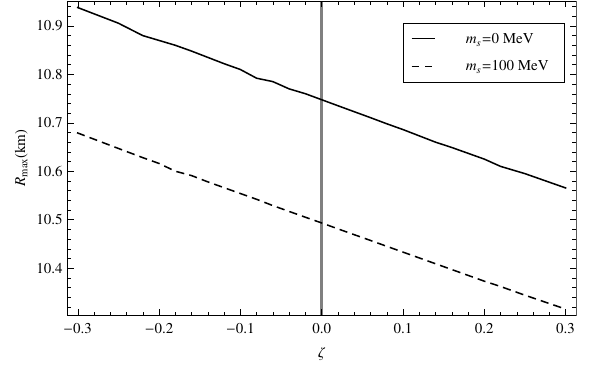}
		\caption{Variation of radius ($R_{max}$) corresponding to the maximum mass with the coupling parameter $\zeta$ for a parametric choice of $n=0.3~fm^{-3}$. The solid and dashed lines represent the variation for $m_s=0~MeV$ and $m_s=100~MeV$, respectively. }\label{fig8}
	\end{minipage}
\end{figure}

\begin{figure}[ht!]
	\centering
	\begin{minipage}{0.45\textwidth}
		\centering
		\includegraphics[width=0.95\linewidth,height=0.65\textwidth]{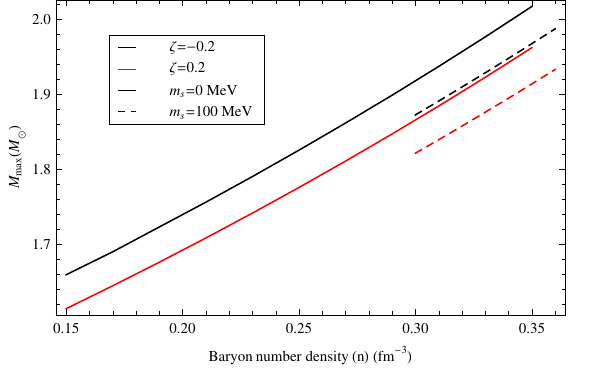}
		\caption{Plot showing the variation of maximum mass ($M_{max}$) with $n$ for different parametric choices of $\zeta$. The black and red line represent the variation for $\zeta=-0.2$ and $0.2$, respectively. Also the solid and dashed lines represent the variation for $m_s=0~MeV$ and $100~MeV$, respectively.}\label{fig9}
	\end{minipage}\hfil
	\begin{minipage}{0.45\textwidth}
		\centering
		\includegraphics[width=0.95\linewidth,height=0.65\textwidth]{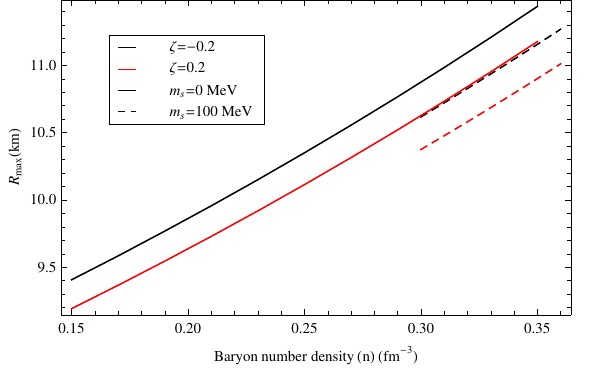}
		\caption{Plot showing the variation of radius corresponding to maximum mass ($R_{max}$) with $n$ for different parametric choices of $\zeta$. The black and red line represent the variation for $\zeta=-0.2$ and $0.2$, respectively. Also the solid and dashed lines represent the variation for $m_s=0~MeV$ and $100~MeV$, respectively.}\label{fig10}
	\end{minipage}
\end{figure}

In this analysis, the Tolman-Oppenheimer-Volkoff (TOV) equations, as presented in references \cite{JRO,RCT}, have been solved numerically to compute the value maximum mass and radius within this framework. To maintain a stable stellar configuration, the values of baryon number density ($n$) are chosen within the stable region as given in Table~\ref{tab1} for each $m_s$ value. In Fig.~\ref{fig3}, the $M-R$ variation for different $n$ values and $\zeta=-0.1$ are shown with a parametric choice of $m_s$. The variation of mass with radius for different $\zeta$ values are shown in Fig.~\ref{fig4} for $n=0.3~fm^{-3}$ and suitable parametric values of $m_s$. Fig.~\ref{fig5} shows the variation of $M$ with the central density for $\zeta=-0.1$ with different $n$ value, and Fig.~\ref{fig6} shows the same variation for $n=0.3~fm^{-3}$ with different $\zeta$ value. In each case, parametric values of $m_s$ are chosen. The dots in each curve in Figs.~\ref{fig3} and \ref{fig4} indicate the maximum mass points maintaining the condition $\frac{\partial M}{\partial \rho_c}>0$ \cite{YBZ2,BKH}. The maximum mass, radius and central density from the model are tabulated in Table~\ref{tab2} and it is evident that with increasing $\zeta$, both the maximum mass ($M_{max}$) and the corresponding radius ($R_{max}$) decrease as gravity matter coupling becomes stronger. In case of GR $(\zeta=0)$, the mass and radius take intermediate values than those for negative and positive value of $\zeta$. Hence, in presence of $\zeta$ results are different than those obtained in GR. We have evaluated the Buchdahl bound $(u_{max})$ in $f(R,T)$ gravity \cite{SB} and it is noted from Table~\ref{tab2} that the modified Buchdahl bound is also satisfied in presence of $\zeta$. In Fig.~\ref{fig7} and \ref{fig8}, the variation of the maximum mass and corresponding radius with the coupling parameter are shown for a parametric choice of $m_s$ and $n=0.3~fm^{-3}$. Notably, both the maximum mass and radius decrease with increasing $\zeta$, and for higher $m_s$ value, this decrease is greater. Now, the variation of $M_{max}$ and $R_{max}$ with baryon number density ($n$) in the stable region are plotted in Figs.~\ref{fig9} and \ref{fig10} for $\zeta=-0.2$ and $0.2$ and a parametric choice of $m_s$. Notably, as $n$ increases $M_{max}$ and $R_{max}$ also increase. 

\begin{table}[ht]
	\centering
	\caption{Table for maximum mass ($M_{max}$), radius ($R_{max}$) and central density ($\rho_c$) and compactness from the TOV solution with parametric choices of $m_s$ and $n$.}\label{tab2}
	\resizebox{0.6\textwidth}{!}{$
		\begin{tabular}{@{}c|c|cccccc}
			\hline
			$m_s$ & $n$        & $\zeta$ & $M_{max}$      & $R_{max}$  & \multicolumn{2}{c}{Compactness}  & $\rho_c\times{10^{15}}$ \\
			$MeV$ & $fm^{-3}$  &         &  $M_{\odot}$ & $km$     & $u=\frac{M_{max}}{R_{max}}$ & $u_{max}$      & $g.cm^{-3}$\\ \hline
			\multirow{6}{*}{0}     & \multirow{3}{*}{0.15}   & -0.1    & 1.65   & 9.35   & 0.2603 & 0.4903 & 2.61 \\ 
			                       &                         & 0.0     & 1.64   & 9.30   & 0.2601 & 0.4938 & 2.66 \\
			                       &                         & 0.1     & 1.62   & 9.24   & 0.2586 & 0.4957 & 2.71 \\ \cline{2-8}  
			                       & \multirow{3}{*}{0.36}   & -0.1    & 2.03   & 11.49  & 0.2606 & 0.4903 & 1.91 \\
			                       &                         & 0.0     & 2.01   & 11.42  & 0.2596 & 0.4938 & 1.98 \\
			                       &                         & 0.1     & 1.99   & 11.35  & 0.2586 & 0.4957 & 2.07\\ \hline
			\multirow{6}{*}{100}   & \multirow{3}{*}{0.3}    & -0.1    & 1.86   & 10.55  & 0.2601 & 0.4903 & 2.11 \\ 
			                       &                         & 0.0     & 1.84   & 10.49  & 0.2587 & 0.4938 & 2.21 \\
			                       &                         & 0.1     & 1.83   & 10.43  & 0.2588 & 0.4957 & 2.25 \\ \cline{2-8}  
			                       & \multirow{3}{*}{0.36}   & -0.1    & 1.98   & 11.20  & 0.2607 & 0.4903 & 2.02 \\
			                       &                         & 0.0     & 1.96   & 11.14  & 0.2595 & 0.4938 & 2.10 \\
			                       &                         & 0.1     & 1.94   & 11.07  & 0.2584 & 0.4957 & 2.16 \\ \hline
		\end{tabular}$}	
\end{table}


\section{Physical application of the present model}\label{sec7}
\begin{figure}[ht!]
	\centering
	\begin{minipage}{0.45\textwidth}
		\centering
		\includegraphics[width=0.85\linewidth,height=0.65\textwidth]{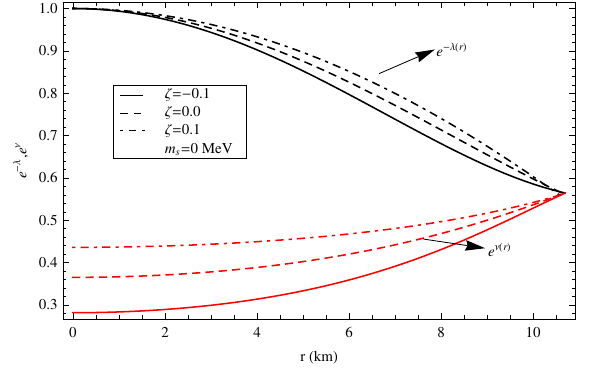}
		\caption{Radial variation of metric potentials inside $4U~1820-30$ for different parametric choices of $\zeta$ taking $m_s=0~MeV$ and $n=0.3~fm^{-3}$. The solid, dashed and dotdashed lines represent value of $\zeta$ as $-0.1$, $0.0$ and $0.1$, respectively. The black and red lines represent the variation of $e^{-\lambda}$ and $e^{\nu}$, respectively.}\label{fig11}
	\end{minipage}\hfil
	\begin{minipage}{0.45\textwidth}
		\centering
		\includegraphics[width=0.85\linewidth,height=0.65\textwidth]{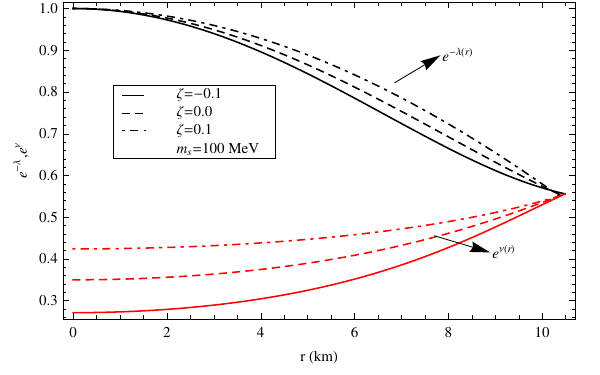}
		\caption{Radial variation of metric potentials inside $4U~1820-30$ for different parametric choices of $\zeta$ taking $m_s=100~MeV$ and $n=0.3~fm^{-3}$. The solid, dashed and dotdashed lines represent value of $\zeta$ respectively as $-0.1$, $0.0$ and $0.1$. The black and red lines represent the variation of $e^{-\lambda}$ and $e^{\nu}$, respectively.} \label{fig12}
	\end{minipage}\hfil
\end{figure}

\begin{figure}[ht!]
	\centering
	\begin{minipage}{0.45\textwidth}
		\centering
		\includegraphics[width=0.95\linewidth,height=0.65\textwidth]{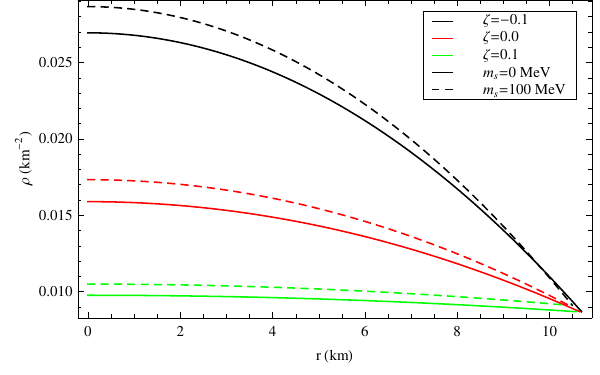}
		\caption{Radial variation of energy density $\rho$ inside $4U~1820-30$. The solid and dashed lines represent $(i)$ $m_s=0~MeV$, $n=0.3~fm^{-3}$ and $(ii)$ $m_s=100~MeV$, $n=0.3~fm^{-3}$, respectively. The blue, red and green lines represent $\zeta=-0.1$, $0.0$ and $0.1$, respectively.}\label{fig13}
	\end{minipage}\hfil
	\begin{minipage}{0.45\textwidth}
		\centering
		\includegraphics[width=0.95\linewidth,height=0.65\textwidth]{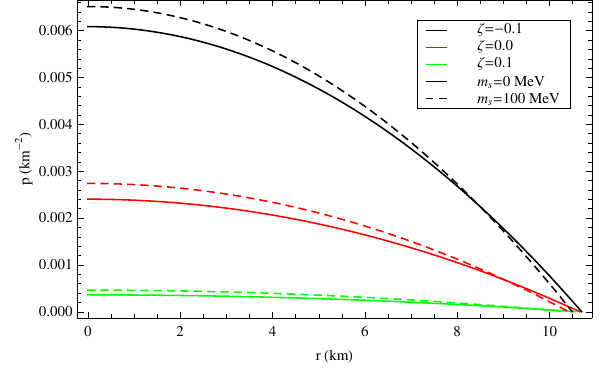}
		\caption{Variation of pressure $(p)$ with $r$ inside $4U~1820-30$. The solid and dashed lines represent $(i)$ $m_s=0~MeV$, $n=0.3~fm^{-3}$ and $(ii)$ $m_s=100~MeV$, $n=0.3~fm^{-3}$, respectively. The blue, red and green lines represent $\zeta=-0.1$, $0.0$ and $0.10$, respectively.}\label{fig14}
	\end{minipage}
\end{figure}

\begin{figure}[ht!]
 	\centering
 	\begin{minipage}{0.45\textwidth}
 		\centering
 		\includegraphics[width=0.95\linewidth,height=0.65\textwidth]{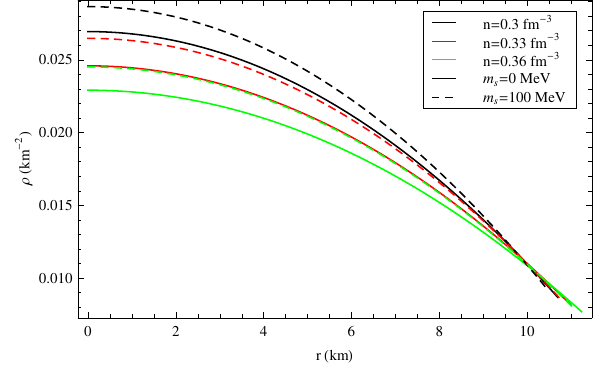}
 		\caption{Radial variation of energy density ($\rho$) inside $4U~1820-30$ for different $n$ taking $\zeta=-0.1$. The solid and dashed lines represent $m_s=0~MeV$ and $m_s=100~MeV$ respectively. The blue, red and green lines represent $n=0.3$, $0.33$ and $0.36~fm^{-3}$  respectively.}\label{fig15}
 	\end{minipage}\hfil
 	\begin{minipage}{0.45\textwidth}
 		\centering
 		\includegraphics[width=0.95\linewidth,height=0.65\textwidth]{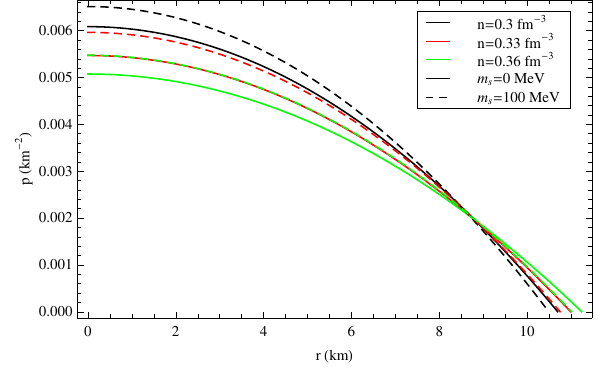}
 		\caption{Variation of pressure $(p)$ with $r$ inside $4U~1820-30$ for different $n$ taking $\zeta=-0.1$. The solid and dashed lines represent $m_s=0~MeV$ and $m_s=100~MeV$ respectively. The blue, red and green lines represent $n=0.3$, $0.33$ and $0.36~fm^{-3}$  respectively.}\label{fig16}
 	\end{minipage}
\end{figure}

\begin{figure}[ht]
	\centering
	\includegraphics{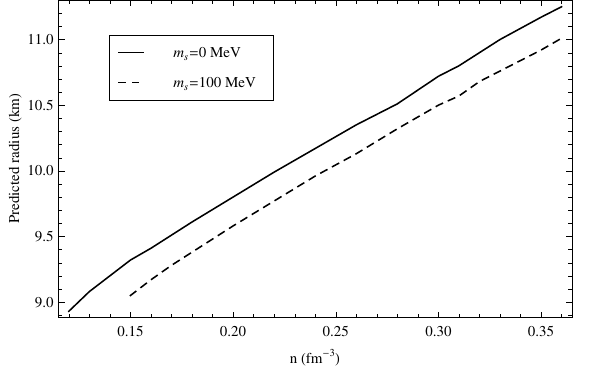}
	\caption{Variation of predicted radius of $4U~1820-30$ with baryon number density ($n$) for a parametric choice of coupling $\zeta=-0.1$. The solid and dashed line represent the variation for $m_s=0~MeV$ and $m_s=100~MeV$, respectively.}\label{fig17}
\end{figure}

This section provides an analysis of the physical properties of a compact object, specifically focusing on its pressure ($p$) energy density ($\rho$) and other relevant properties in $f(R,T)$ gravity. In this analysis, the effect of baryon number density $(n)$ on such variables in the context of $f(R,T)$ gravity for a finite strange quark mass ($m_s$) is studied. For physical inspection, the X-ray binary $4U~1820-30$ with an observed mass $1.58~M_{\odot}$ \cite{TGA} is considered. Variation of the $g_{rr}$ and $g_{tt}$ metric components for the selected compact object are shown in Figs.~\ref{fig11} and \ref{fig12} for suitable parametric choice of model parameters. The variation of $\rho$ and $p$ inside the compact object are shown in Figs.~\ref{fig13} and \ref{fig14}, respectively, for various parametric choice of model parameters. The plots show that the density and pressure trend are continuous and in accordance with realistic properties. It is evident from Figs.~\ref{fig13} and \ref{fig14} that an increase in coupling ($\zeta$) decreases both $\rho$ and $p$ for a parametric choice of $m_s$ and $n$. The variation of $\rho$ and $p$ for different $n$ values within stable region are shown in Figs.~\ref{fig15} and \ref{fig16}. It is evident that for higher $n$ value pressure support decreases and hence radius increases. As a physical plausibility of the study, the radii of several compact objects are predicted within the allowed parameter space of the model and the results are tabulated in Table~\ref{tab3}. The variation of predicted radius for the chosen compact object $4U~1820-30$ with $n$ is shown in Fig.~\ref{fig17} for suitable parametric choice of $m_s$ and $\zeta$.

\begin{table}[ht]
	\centering
	\caption{Prediction of the radius of several compact objects within the allowed parameter space}\label{tab3}
	\resizebox{0.9\textwidth}{!}{$
		\begin{tabular}{@{}c|cc|ccc|ccc}
				\hline
		Name of the      & Mass             & Observed        & \multicolumn{6}{c}{Predicted radius}\\ \cline{4-9}
		compact object   & ($M_{\odot}$)    & radius          & \multicolumn{3}{c}{$m_s=0~MeV$} \vline &   \multicolumn{3}{c}{$m_s=100~MeV$}\\ \cline{4-9}
		                 &                  &  ($km$)         & $\zeta$ & $n~(fm^{-3})$& $R~(km)$  & $\zeta$ & $n~(fm^{-3})$ &  $R~(km)$ \\ \hline  
		$VELA~X-1$       & 1.77\cite{TGA}   & 9.56$\pm$0.08   &  0.18   & 0.25  & 9.87 & 0.15   & 0.30 & 10.33 \\
		$4U~1608-52$     & 1.74\cite{TGU1}  & 9.3$\pm$1.0     &  0.16   & 0.24  & 9.88 & 0.14   & 0.30 & 10.38\\
		$4U~1820-30$     & 1.58\cite{TGU2}  & 9.11$\pm$0.4    &  0.14   & 0.15  & 9.14 & 0.42   & 0.30 & 10.39\\
		$PSR~J0030+451$  & 1.44\cite{MCM}   & 13.02           & -0.3    & 0.36  & 11.16& -0.3   & 0.36 & 10.93\\
		$SMC~X-4$        & 1.29\cite{TGA}   & 8.831$\pm$0.09  & 0.08    & 0.15  & 9.09 & 0.05   & 0.30 & 10.03 \\
		$LMC~X-4$        & 1.04\cite{TGA}   & 8.301$\pm$0.2   & 0.05    & 0.15  & 8.71 & 0.04   & 0.30 & 9.56\\
		$Her~X-1$        & 0.85\cite{TGA}   & 8.1$\pm$0.41    & 0.04    & 0.15  & 8.25 & 0.04   & 0.30 & 8.98\\ \hline
		\end{tabular}$}
\end{table}

\subsection{Causality condition}\label{sec7.1}
For a viable stellar model, the causality condition must hold at all internal points and at the surface. For an isotropic case, the expression of the sound velocity is $v^2=\frac{dp}{d\rho}$. Now, thermodynamic stability and causality restrict the velocity of sound to $0\leq v^2\leq{1}$. Now, using the equations for $p$ and $\rho$ as given in Eqs.~(\ref{10}) and (\ref{21}), the expression for the sound velocity in this model is $v^2=\frac{dp}{d\rho}=\frac{1}{3}$. So the causality condition holds good for the parametric choices of $\zeta$, $n$ and $m_s$ used to construct the stellar model.

\subsection{Energy condition}\label{sec7.2}
\begin{figure}[ht]
	\centering
	\begin{minipage}{0.45\textwidth}
		\centering
		\includegraphics[width=0.95\textwidth,height=0.75\linewidth]{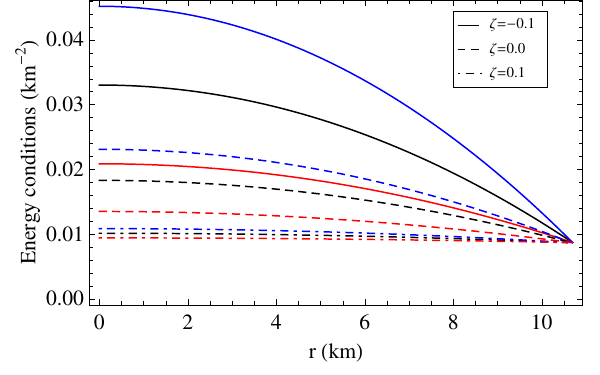}
		\caption{Variation of different energy conditions inside $4U~1820-30$ for $m_s=0~MeV$ and $n=0.3~fm^{-3}$. The solid, dashed and dotdashed lines represent the variation for $\zeta=-0.1$, $0.0$ and $0.1$, respectively. The black, blue and red lines represent the variation of $\rho+p$, $\rho+3p$ and $\rho-p$, respectively. }\label{fig18}
	\end{minipage}\hfil
	\begin{minipage}{0.45\textwidth}
		\centering
		\includegraphics[width=0.95\textwidth,height=0.75\linewidth]{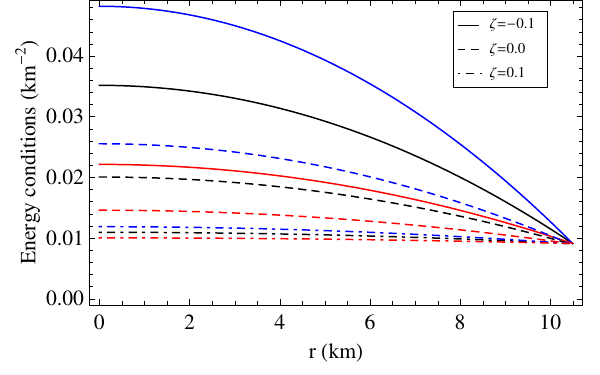}
		\caption{Variation of different energy conditions inside $4U~1820-30$ for $m_s=100~MeV$ and $n=0.3~fm^{-3}$. The solid, dashed and dotdashed lines represent the variation for $\zeta=-0.1$, $0.0$ and $0.1$, respectively. The black, blue and red lines represent the variation of $\rho+p$, $\rho+3p$ and $\rho-p$, respectively.}\label{fig19}
	\end{minipage}		
\end{figure}

Although the curvature of space-time and matter content inside a compact object is related via Eq.~(\ref{15}) in the $f(R,T)$ gravity model, there is ample amount of arbitrariness in the specific form of $T_{\sigma\eta}$. Therefore, constraints must be imposed on $T_{\sigma\eta}$. In gravitational theory, the energy conditions ensure the physical plausibility of the energy-momentum tensor. The characteristics of internal matter distribution without requiring an explicit definition of matter content, can be analysed by imposing such conditions. Additionally, without directly relying on specific expressions for pressure or energy density, it becomes feasible to extract insights into extreme phenomena, such as gravitational collapse. In mathematical terms, analysing the energy conditions involves addressing the problem of eigenvalue associated with the fluid energy-momentum tensor \cite{CAK}. In a $4$-dimensional spacetime, this leads to solving a quartic polynomial equation to determine the eigenvalues, which can be challenging owing to the lack of straightforward analytical solutions. Nonetheless, for a physically meaningful and realistic matter distribution, the four energy conditions namely, ($i$) Null (NEC), ($ii$) Strong (SEC), ($iii$) Dominant (DEC) and ($iv$) Weak (WEC) energy conditions must be satisfied within the boundary of the stellar configuration \cite{CAK,SWH,RWA}. The mentioned energy conditions for an isotropic fluid sphere can be expressed mathematically as \cite{BP,BPB}:
\begin{enumerate}
	\item Null Energy Condition (NEC: $\rho + p\geq 0$).
	\item Strong Energy Condition (SEC: $\rho + p\geq 0,~\rho + 3p\geq 0$).
	\item Dominant Energy Condition (DEC: $\rho\geq 0,~\rho - p\geq 0$).
	\item Weak Energy Condition (WEC: $\rho + p\geq 0,~\rho\geq0 $).
\end{enumerate} 
The radial variations of different energy conditions within the compact object $4U~1820-30$ are shown in Figs.~\ref{fig18} and \ref{fig19} for different parametric combinations of the model parameters. Notably, all the energy conditions are satisfied from centre to the surface within this choice of parameter space. 

\section{Stability analysis}\label{sec8}
The following stability check has been performed to verify the validity of the model: 
\begin{enumerate}
	\item Hydrostatic equilibrium under the influence of different forces
	\item Calculation of the adiabatic index
	\item Stability under Lagrangian perturbation of the absolute value of pressure
	\item Tidal deformability and Tidal love number determination
\end{enumerate}
\subsection{Hydrostatic equilibrium under the joint action of different forces}\label{sec8.1}
\begin{figure}[ht]
	\centering
	\begin{minipage}{0.45\textwidth}
		\centering
		\includegraphics[width=0.95\textwidth,height=0.75\linewidth]{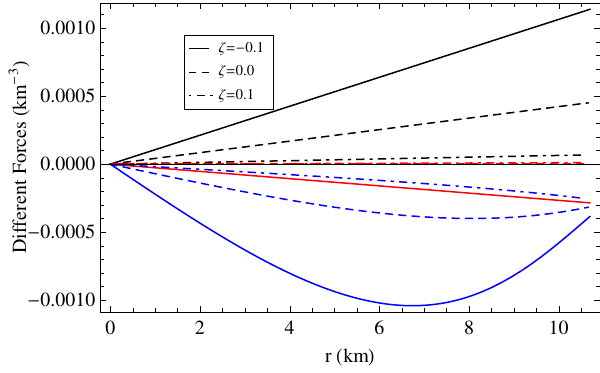}
		\caption{Variation of different forces inside $4U~1820-30$ with $r$ for $n=0.3~fm^{-3}$ and $m_s=0~MeV$. The black, blue and red lines represent the gravity force ($F_g$), hydrostatic force ($F_h$) and force due to modified gravity ($F_\zeta$), respectively. Here the solid, dashed and dotdashed lines represent values of the coupling parameter as $\zeta=-0.1$, $0.0$ and $0.1$, respectively.}\label{fig20}
	\end{minipage}\hfil
	\begin{minipage}{0.45\textwidth}
		\centering
		\includegraphics[width=0.95\textwidth,height=0.75\linewidth]{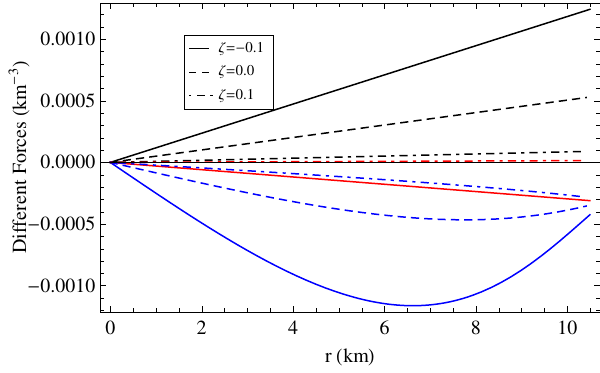}
		\caption{Radial variation of different forces inside $4U~1820-30$ for $n=0.3~fm^{-3}$ and $m_s=100~MeV$. The black, blue and red lines represent the gravity force ($F_g$), hydrostatic force ($F_h$) and force due to modified gravity ($F_\zeta$), respectively. Here the solid, dashed and dotdashed lines represent values of the coupling parameter as $\zeta=-0.1$, $0.0$ and $0.1$, respectively.}\label{fig21}
	\end{minipage}
\end{figure}

In presence of different forces inside the stellar configuration fluid should follow the stable hydrostatic equilibrium condition. In the present case, the equilibrium between all forces can be discussed by the following generalised TOV equation \cite{JRO,RCT} as given below:
\begin{equation}
	-\frac{M_{G}(r)(\rho+p)}{r^2}e^{(\lambda-\nu)/2}-\frac{dp}{dr}+\frac{\zeta}{1+2\zeta}\Big({p}^{\prime}-{\rho}^{\prime}\Big)=0,\label{32}
\end{equation}
$M_{G}(r)$ is the active gravitational mass contained inside a spherical region of radius $r$ and can be obtained from the formula of Tolman-Whittaker \cite{OGR} and using the EFE as:
\begin{equation}
	M_{G}(r)=\frac{1}{2}r^{2}\nu^{\prime}e^{(\nu-\lambda)/2}.\label{33}
\end{equation}
Substituting Eq.~(\ref{33}) into (\ref{32}), the final form of the expression becomes:
\begin{equation}
	-\frac{\nu^{\prime}}{2}(\rho+p)-\frac{dp}{dr}+\frac{\zeta}{1+2\zeta}({p}^{\prime}-{\rho}^{\prime})=0.\label{34}
\end{equation} 
The third term in the right hand side of Eq.~(\ref{34}) arises because of the fact that the energy-momentum tensor is not conserved in $f(R,T)$ gravity. The extra force breaks the balance between gravity and pressure alone. There is an additional effective force acting on fluid elements due to the geometry–matter coupling. The pressure gradient required to counteract gravity is altered in the present scenario. For an arbitrary value of $\zeta$, this extra force often reduces the effective gravitational pull, making the star more supportive against collapse. However, this extra force has direct effect on tidal deformability, gravitational redshift and compactness. The equilibrium inside an isotropic star under the combined effect of three forces is represented by the condition given in Eq.~(\ref{34}). These forces are, force due to gravity ($F_{g}$), hydrostatic force denoted by ($F_{h}$) and force due to modified gravity ($F_\zeta$) where, $F_g=-\frac{\nu^{\prime}}{2}(\rho+p)$, $F_h=-\frac{dp}{dr}$ and $F_\zeta=\frac{\zeta}{1+2\zeta}({p}^{\prime}-{\rho}^{\prime})$. Eq.~(\ref{34}) yields that the net force $F=F_g+F_h+F_\zeta$ should be zero inside a star. The radial variations of different forces inside $4U~1820-30$ are shown in Figs.~\ref{fig20} and \ref{fig21} for different parametric combinations of $m_s$, $n$ and the coupling parameter $\zeta$. Notably, the model maintains hydrostatic equilibrium under the influence of different forces.

\subsection{Calculation of the adiabatic index}\label{sec8.2}
\begin{figure}[ht]
	\centering
	\begin{minipage}{0.45\textwidth}
		\centering
		\includegraphics[width=0.9\textwidth,height=0.75\linewidth]{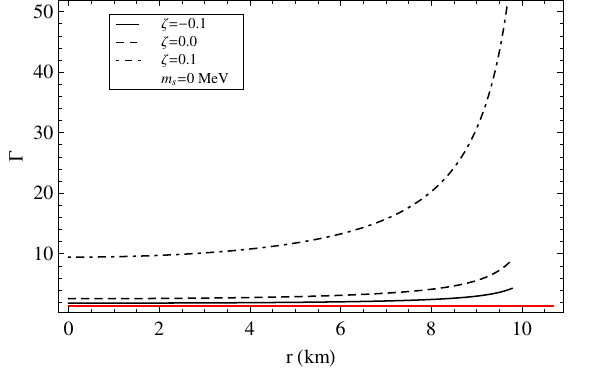}
		\caption{Variation of the adiabatic index ($\Gamma$) with $r$ inside $4U~1820-30$ for $n=0.3~fm^{-3}$ and $m_s=0~MeV$. The solid, dashed and dotdashed lines represent the variation for $\zeta=-0.1$, $0.0$ and $0.1$, respectively. The solid red line represents the Newtonian (isotropic) limit of $\Gamma=\frac{4}{3}$}\label{fig22}
	\end{minipage}\hfil
	\begin{minipage}{0.45\textwidth}
		\centering
		\includegraphics[width=0.9\textwidth,height=0.75\linewidth]{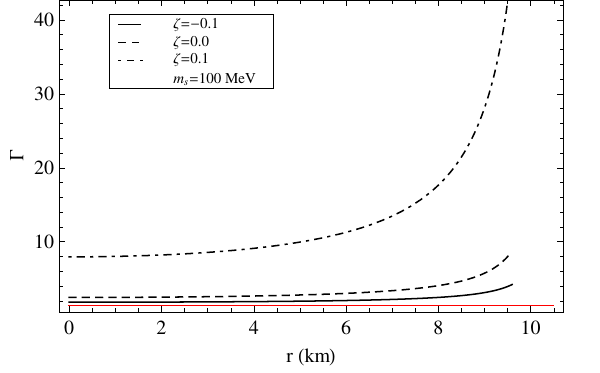}
		\caption{Radial variation of the adiabatic index ($\Gamma$) inside $4U~1820-30$ for $n=0.3~fm^{-3}$ and $m_s=100~MeV$. The solid, dashed and dotdashed lines represent the variation for $\zeta=-0.1$, $0.0$ and $0.1$, respectively. The solid red line represents the Newtonian (isotropic) limit of $\Gamma=\frac{4}{3}$}\label{fig23}
	\end{minipage}
\end{figure}

The adiabatic index of fluid inside a star plays an important role in determining the condition of stable stellar structure. The adiabatic index ($\Gamma$), in general, of a fluid sphere describes how effectively counter resistive force originated inside a fluid on being compressed. Heintzmann and Hillebrandt \cite{HHE} considering relativistic approach have shown that in case of a Newtonian perfect fluid of isotropic pressure, the value of adiabatic index is $\frac{4}{3}$ and for any isotropic stable stellar structure, $\Gamma>\frac{4}{3}$ yields positive normal mode frequencies. Hence, to ensure stability, $\Gamma>\frac{4}{3}$ condition must be satisfied. In the $f(R,T)$ theory of gravity, the mathematical expression for $\Gamma$ related to compact star can be written as:
\begin{equation}
	\Gamma=\frac{\rho+p}{p}\left(\frac{dp}{d\rho}\right). \label{35}
\end{equation}
Figs.~\ref{fig22} and \ref{fig23}, clearly show that the model is highly stable with respect to the radial variation of the adiabatic index $\Gamma$ for the parameter space used here. 
\subsection{Stability under Lagrangian perturbation of the absolute value of pressure}\label{sec8.3}
\begin{figure}[ht!]
	\centering
		\includegraphics{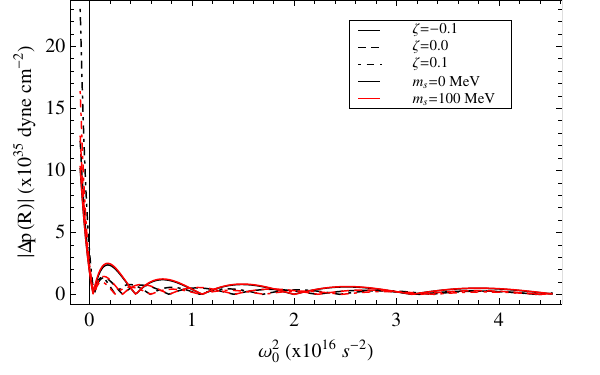}
		\caption{Variation of absolute value of pressure ($|\Delta p(R)|$) with normal mode frequencies ($\omega_0^2$) inside $4U~1820-30$ for $n=0.3~fm^{-3}$. The black and red lines represent $m_s=0~MeV$ and $100~MeV$, respectively. Different value of $\zeta$ as $\zeta=-0.1$, $0.0$ and $-0.1$ is represented by the solid, dashed and dotdashed lines, respectively.}\label{fig24}
\end{figure}

The stability of a compact star can be studied within the context of $f(R,T)$ gravity. For such study it is necessary to evaluate the frequencies of the normal mode of oscillations. In this context, a small deviation from the condition of hydrostatic equilibrium is considered. A fluid element in this approach located at the position $r$ is displaced to $r+\xi(x^0,r)$ in the presence of perturbation, without losing spherical symmetry. The perturbation will cause motion in the radial direction. Now, the equations governing such radial perturbation in pressure within the context of modified gravity theory $f(R,T)=R+2\zeta T$ is \cite{JMZ}:
\begin{equation}
	\frac{d\xi}{dr}=-\frac{1}{r}\Big[3\xi +\frac{\mathcal{J}}{\Gamma p}\Delta p\Big]+\nu^{\prime}\xi,\label{36}
\end{equation}
\begin{eqnarray}
	\mathcal{G}\frac{d\Delta p}{dr}=\xi\Bigg[\frac{1+a}{1+2a}(\rho+p)\Big[\frac{\omega_0^2}{c^2}e^{-\nu}-(k+3\zeta)p+\zeta\rho\Big]re^{\lambda}+\frac{1-a}{1+2a}(\rho+p)r{\nu^{\prime}}^2-\frac{1+3a}{1+2a}4{p}^{\prime}\nonumber\\ \frac{ar}{1+2a}\Big[\frac{4{\rho}^{\prime}}{r}+a({\rho}^{\prime}-{p}^{\prime})\Big(\frac{9}{r}+\nu^{\prime}+\lambda^{\prime}\Big){p}^{\prime}\Big(\frac{1}{r}+\nu^{\prime}+\lambda^{\prime}\Big)-(\rho+p)\Big(\lambda^{\prime}+\frac{2}{r}\Big)\nu^{\prime}\Big]\Bigg]\nonumber\\ -\frac{ar}{1+2a}\Big[a({\rho}^{\prime}-{p}^{\prime})-{p}^{\prime}\Big]\xi^{\prime}-\Delta p\Biggl\{2\nu^{\prime}+\lambda^{\prime}-\frac{a}{1+2a}\Big[(\lambda^{\prime}+3\nu^{\prime})\frac{\rho+p}{\Gamma p}\mathcal{J}\frac{d}{dr}\Big(\frac{\rho+p}{\Gamma p}\mathcal{J}\Big)\Big]\Biggr\},\label{37}
\end{eqnarray}
where, $\mathcal{J}=\frac{k+3\zeta}{k+2\zeta}\Big(1-\frac{\zeta}{k+3\zeta}\frac{dp}{d\rho}\Big)$, $\mathcal{G}=1-\frac{a}{1+2a}\Big(\frac{\rho+p}{\Gamma p}\mathcal{J}\Big)$, $a=\frac{\zeta}{k+2\zeta}$ and $k=\frac{8\pi G}{c^4}=1$. The values of metric potentials $\lambda$ and $\nu$ are obtained from Eqs.~(\ref{22}) and (\ref{23}), respectively. To solve the system of coupled Eqs.~(\ref{36}) and (\ref{37}) the procedure adopted by Pretel et al. \cite{JMZ} has been used in this model. Variation of absolute value of Lagrangian perturbation in pressure ($\Delta p$) with eigenfrequencies ($\omega_0^2$) inside $4U~1820-30$ is shown in Fig.~\ref{fig24} for different parametric combinations of $\zeta$, $m_s$ and $n=0.3~fm^{-3}$. The plots show that all the eigenfrequencies are positive, which indicates the stability of the model against small radial perturbation.  
\subsection{Tidal deformability and Tidal love number}\label{8.4}
\begin{figure}[ht]
	\centering
	\begin{minipage}{0.45\textwidth}
		\centering
		\includegraphics[width=0.9\textwidth,height=0.75\linewidth]{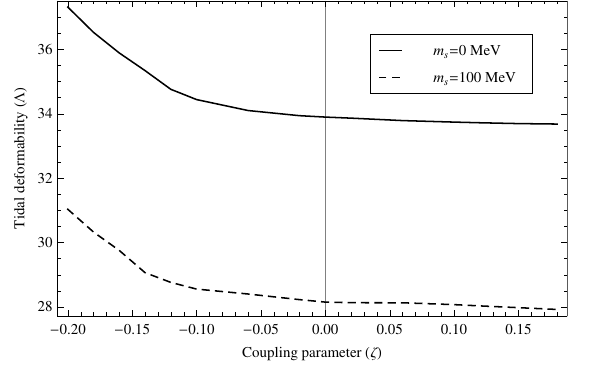}
		\caption{Variation of tidal deformability ($\Lambda$) with coupling parameter ($\zeta$) for the compact object $4U~1820-30$ with a parametric choice of $n=0.3~fm^{-3}$. The solid and dashed lines represent the variation for $m_s=~0$ and $100~MeV$, respectively. }\label{fig25}
	\end{minipage}\hfil
	\begin{minipage}{0.45\textwidth}
		\centering
		\includegraphics[width=0.9\textwidth,height=0.7\linewidth]{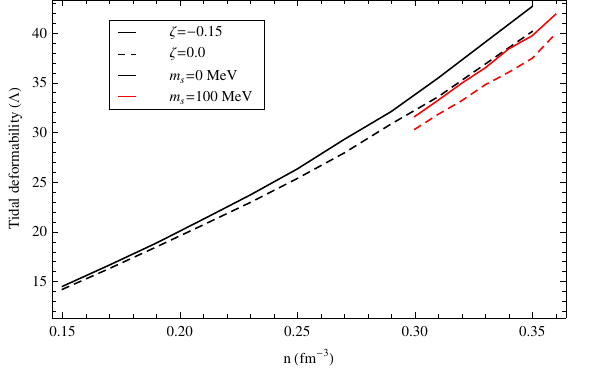}
		\caption{Variation of tidal deformability ($\Lambda$) with $n$ for the compact object $4U~1820-30$ for parametric choices of $\zeta$. The solid and dashed lines represent the variation for $\zeta=-0.2$ and $0.0$, respectively. The black and red lines represent the variation for $m_s=0~MeV$ and $100~MeV$, respectively.}\label{fig26}
	\end{minipage}
\end{figure}

Tidal deformability refers to the extent to which a compact object, such as a neutron star, deforms in response to the field of a companion body. This deformation is caused by tidal forces and depends on the internal structure of the object. The Tidal Love Number (TLN) quantifies this deformation and represents the object's resistance to external tidal effects. It is a dimensionless parameter that characterizes how the mass distribution within the body changes due to the external field. Owing to the deformation caused by this external field $\epsilon_{ij}$, a quadrupole moment $Q_{ij}$ is developed inside a compact object. The tidal deformability parameter is then defined as $\lambda_{tidal-def}=-\frac{Q_{ij}}{\epsilon_{ij}}$. Additionally, the dimensionless deformability parameter is given as $\Lambda=\frac{\lambda_{tidal-def}}{M^5}$, where, $M$ is the mass of the compact object. Furthermore, the tidal deformability ($\Lambda$) is connected to the tidal Love number $k_2$ via the following specific relation \cite{EEF}:
\begin{equation}
	k_2=\frac{3}{2}\Lambda\frac{M}{R^5}.\label{38}
\end{equation}
where $R$ is the radius of the compact object. Following the method used in the articles \cite{KBG,KST,TRE}, the behaviour of the fluid configuration in equilibrium under linearised perturbations due to the presence of an external tidal field for the $l=2$, (even-parity) in modified $f(R,T)$ gravity theory can be analysed. The background metric $g_{ij}$ is perturbed in the presence of an external tidal field and represented by the following modified metric:
\begin{equation}
	\tilde{g}_{ij}=g_{ij}+h_{ij},\label{39}
\end{equation} 
where, $h_{ij}$ is a linearised metric perturbation. Following the methods of previous studies \cite{KST,TRE} $h_{ij}$ can be expressed as:
\begin{equation}
h_{ij}=diag[H_0e^\nu,H_2(r)e^\lambda,r^2K(r), r^2\sin 2\theta\;\;K(r)]Y_{2m}(\theta,\phi).\label{40}
\end{equation}
Now, according to the work of Goswami et al. \cite{KBG}, the equation for $H(r)$ in $f(R,T)=R+2\zeta T$ is given as:
\begin{equation}
	H^{\prime\prime}(r)+f_1 H^{\prime}(r)+f_2 H(r)=0,\label{41}
\end{equation}
where, 
\begin{equation}
	f_1=\frac{1}{r}\Big[1+e^{\lambda}\Big(1+2r^2p(0.25+\zeta)-\frac{r^2\rho}{2}-2r^2\zeta\rho\Big)\Big].\label{42}
\end{equation}  
\begin{eqnarray}
	f_2=-\frac{1+4e^{\lambda}+e^{2\lambda}}{r^2}+e^{\lambda}\left(\frac{1}{2}+\zeta\right)\left(p+\rho\right)+\frac{e^{\lambda}(0.5+\zeta)\Big[1-\zeta(v^2-3)\Big]\Big(p+\rho\Big)}{v^2+(3v^2-1)\zeta}\nonumber \\ -e^{2\lambda}r^2\Big[p+\zeta\left(3p-\rho\right)\Big]^2-2e^{\lambda}\Big[p(e^{\lambda}-3)+\zeta p(3e^{\lambda}-8)-(1+\zeta e^{\lambda})\rho\Big].\label{43}
\end{eqnarray}
and $v^2=\frac{dp}{d\rho}$ is the velocity of sound. Again following the work of Ref.~\cite{THI,THI2} the tidal love number ($k_2$) is given in the following form:
\begin{eqnarray}
	k_2=\frac{8u^5}{5}\left(1-2u\right)^2\Bigg(2+2u(y-1)-y\Bigg)\Big[2u\{6-3y+3u(5y-8)\}+4u^3\{13-11y+u(3y-2)+2u^2(1+y)\}\nonumber\\ +3(1-2u)^2\{2-y+2u(y-1)\log(1-2u)\}\Big]^{-1},\label{44}
\end{eqnarray}
where, $u=\frac{M}{R}$ and $y=\frac{R H^{\prime}(R)}{H(R)}$, where $H^{\prime}(R)$ is the value of $H(r)$ at the surface ($r=R$). On the basis of data from the merger of binary neutron star $GW~170817$, Abbott and Abbott \cite{BPA} established constraints on the parameter $\Lambda$, which offers key insight into the NS EoS. According to Bauswein et al \cite{ABA}, the tidal deformability for an NS with a mass of $1.4~M_{\odot}$ should be less than $800$. In Table~\ref{tab4}, the values of the tidal love number ($k_2$) and tidal deformability ($\Lambda$) are tabulated for various compact objects with the parameter space as noted in Table~\ref{tab3}. The variation of $\Lambda$ with coupling parameter $\zeta$ for $4U~1820-30$ for parametric choices of $m_s$ is shown in Fig.~\ref{fig25}. Notably, $\Lambda$ decrease with increasing coupling parameter, which may be attributed to the fact that with increasing $\zeta$, the impact of coupling increases and consequently the resistivity to external deformation increases. As a result $\Lambda$ decrease. Again, with increasing $m_s$ the value of $\Lambda$ decrease than compared to the case for $m_s=0~MeV$ for a given $\zeta$. Also the variation of $\Lambda$ with baryon number density ($n$) in the stable region is shown in Fig.~\ref{fig26} for different parametric combinations of model parameters. It is evident from Fig.~\ref{fig25} that increasing $n$ within the star also increases the tidal deformability.  

\begin{table}[ht]
	\centering
	\caption{Tidal love number and tidal deformability for the selected compact objects with parameters as mentioned in Table~\ref{tab3}}\label{tab4}
	\resizebox{1.0\textwidth}{!}{$
		\begin{tabular}{@{}c|c|ccc|ccc}
			\hline
		Name of the & Observed Mass & \multicolumn{3}{c}{$m_s=~0~MeV$} \vline& \multicolumn{3}{c}{$m_s=~100~MeV$}\\ \cline{3-8}
		compact object& $(M_{\odot})$& $R~(km)$ & $k_2$ & $\Lambda$& $R~(km)$ & $k_2$ & $\Lambda$\\ \hline 
		$VELA~X-1$       & 1.77\cite{TGA}   & 9.87      & 0.0231 & 11.8797 & 10.33  & 0.0240 & 15.4893\\
		$4U~1608-52$     & 1.74\cite{TGU1}  & 9.88      & 0.0233 & 13.1575 & 10.38 & 0.02437 & 17.5786\\
		$4U~1820-30$     & 1.58\cite{TGU2}  & 9.14      & 0.0236 & 14.5973 & 10.39 & 0.0300 & 35.2281\\
		$SMC~X-4$        & 1.29\cite{TGA}   & 9.09      & 0.0264 & 43.8278 & 10.03 & 0.0269 & 73.1258\\
		$PSR~J0030+451$  & 1.44\cite{MCM}   & 11.16     & 0.0369 & 98.4857 & 10.93 & 0.0357 & 85.7732\\
		$LMC~X-4$        & 1.04\cite{TGA}   & 8.71      & 0.0272 & 106.8540 & 9.56  & 0.0270 & 169.5060\\
		$HER~X-1$        & 0.85\cite{TGA}   & 8.25      & 0.0268 & 220.1750 & 8.98 & 0.0262 & 329.376\\ \hline
 		\end{tabular}$}
\end{table}

\section{Discussion}\label{sec9}
This study examines the impact of a nonzero strange quark mass ($m_s$) on various physical parameters in the context of modified $f(R,T)=R+2\zeta T$ gravity theory, where $\zeta$ is the coupling parameter. The internal matter is considered to be composed of deconfined quarks ($u$, $d$ and $s$) and electrons so that the overall charge neutrality condition is maintained. The dynamics of quark confinement between these quarks may be described by MIT bag model. The structure of compact stars is strongly influenced by the choice of the EoS, particularly in the high-density regime. In the context of strange star (SS) modeling, the MIT bag model serves as a fundamental framework for describing the deconfined quark phase at extreme densities. Experimental findings from CERN have firmly established that at densities of approximately $10^{15}~gm/cm^{3}$, transition from hadronic matter phase to a phase of quark-gluon plasma is likely to occur. However, the conventional MIT bag model, with constant $B_g$, may not fully capture the characteristics of such phase transition. To construct a more physically realistic EoS, a baryon number density-dependent bag function $B(n)$ is introduced, where $B(n)$ varies with the baryon number density $n$ as formulated in the work of Liu et al. \cite{YXL} and parametrised by Prasad and Bhalerao \cite{NPR}, which is given in Eq.~(\ref{11}). To check for stability, the energy per baryon ($E_B$) is calculated for different parametric choices of $m_s$. As noted in Fig.~\ref{fig1} with increasing $n$, $E_B$ decreases and this falling nature is more steeper for $m_s=120~MeV$ than $m_s=0~MeV$. Thus, regarding the energy per baryon value, the range of $n$ is restricted within the constraint that the bag value should be within the range of $57.55~MeV/fm^3$ to $95.11~MeV/fm^3$ for different $m_s$ values. Also depending on the value of $E_B$, three different stability windows are obtained. All the results are tabulated in Table~\ref{tab1}. Notably, with increasing $m_s$, the range of $n$ for stable window ($E_B\leq 930.4~MeV$) decreases, with the maximum value being $n=0.36~fm^{-3}$, irrespective of $m_s$. In Fig.~\ref{fig2}, the variation of different chemical potentials against baryon number density ($n$) is shown with a parametric value of $m_s$. Although $\mu$ decreases with $n$, $\mu_e$ slightly increases for the situation $m_s\neq 0$. This change in $\mu$ and $\mu_e$ is more prominent for higher $m_s$ value. The chemical potential can decrease with increasing baryon number density under specific physical conditions such as $(i)$ Strong attractive interactions dominate (e.g., nuclear saturation) and $(ii)$ Phase transitions or mixed phases allow changes in density at constant or decreasing $\mu$. To investigate the system in modified gravity, the density function $\rho$ is taken as given in Eq.~(\ref{21}), which reduces to form formulated by Mak and Harko \cite{MH} for $\zeta=0$. Now, using the EoS for the system given in Eq.~(\ref{10}) along with this choice of $\rho$, as given in Eq.~(\ref{21}), the solutions for various physical parameters are obtained and given in Eqs.~(\ref{22})-(\ref{27}) respectively. To find a constraint on the coupling parameter $\zeta$, it is assumed that the central energy density should be positive and higher than the surface density. Implementing these conditions, the range of $\zeta$ is obtained, which is given in Eqs.~(\ref{30}) and (\ref{31}), respectively. Notably, the range of $\zeta$ is highly mass radius specific. The maximum mass ($M_{max}$) and corresponding radius ($R_{max}$) are obtained through numerical solutions of the modified TOV equations for hydrostatic equilibrium for different parametric combinations of $m_s$, $\zeta$ and $n$ values within the stable region of quarks. Figs.~\ref{fig3} and \ref{fig4} show the variations of mass with radius for $(i)$ $\zeta=-0.1$ and different $n$ values and $(ii)$ for $n=0.3~fm^{-3}$ and suitable $\zeta$ values. For both cases, different $m_s$ values are chosen for study. The variation of mass with central density are shown in Figs.~\ref{fig5} and \ref{fig6}. The black dots on each curve represent the points up to which the static stability criterion $(\frac{\partial M}{\partial{\rho_c}}>0)$ as given by Harrison-Zeldovich-Novikov \cite{YBZ2,BKH} is satisfied. All the outcomes are tabulated in Table~\ref{tab2}. The variation of maximum mass ($M_{max}$) and corresponding radius ($R_{max}$) with $\zeta$ for $n=0.3~fm^{-3}$ and with $n$ for parametric choice of $\zeta$ are shown in Figs.~\ref{fig7}-\ref{fig10}, respectively. These plots indicate that as the coupling increases, both $M_{max}$ and $R_{max}$ decrease, whereas with increasing $n$ value they increase for different parametric choices of $m_s$. Therefore, to include higher mass limits, a smaller value of $\zeta$ with a high value of number density is preferable. \par 

As a physical analysis of the model, the radii of different compact objects with known masses, which are assumed to be strange stars, are predicted, and the results are listed in Table~\ref{tab3}. To show the effects of $\zeta$ and $n$ on various physical parameters, the compact object $4U~1820-30$ with a mass of $1.58~M_{\odot}$ and an observed radius of $9.1\pm 0.4~km$ \cite{TGA} is chosen. The radial variation of metric potentials, density ($\rho$), pressure ($p$) for the chosen compact star are shown in Figs.~\ref{fig11}-\ref{fig14}, respectively. It is evident from these plots that all the parameters are within the allowed range and satisfies all the necessary boundary conditions. From Fig.~\ref{fig11}, it is noted that with increasing $\zeta$, although $e^{-\lambda}$ remains the same near the centre, $e^{\nu}$ increases as $r\rightarrow0$. The same kind of variation in metric potential is observed for $m_s\neq0$, as shown in Fig.\ref{fig12}. As the metric potentials are directly related to the total mass within a radius, it is concluded that gravity matter coupling should have some effect on these parameters. Figs.~\ref{fig13} and \ref{fig14}, show that with increasing $\zeta$ central value of density ($\rho$), pressure ($p$) decrease. In Figs.~\ref{fig15} and \ref{fig16} radial variation of energy density ($\rho$) and pressure ($p$) is shown for parametric choice of $\zeta$ and $m_s$, taking different $n$ values. Notably, both the central values of $\rho$ and $p$ decrease with increasing $n$. Although this decrease is slightly higher for $m_s=0~MeV$ than $m_s=100~MeV$. It is also evident from Fig.~\ref{fig16} that radius increases with $n$ and this increment is higher for $m_s=0~MeV$ than $m_s=100~MeV$. The variation of predicted radius from the model for $4U~1820-30$ is shown in Fig.~\ref{fig17} for the parametric choice of model parameters. From Fig~\ref{fig17}, it is interesting to note that when mass of the strange quark ($m_s$) increases, predicted radius from the model also decreases for a fixed value of $n$. However, radius predicted from the model may be comparable to the radius estimated from observations for lower value of $n$. But lower value of $n$ corresponds to unstability through metastable state as tabulated in Table~\ref{tab1}. Such results corresponds to the possibility of formation of a mixed phase (quark and hadrons or hybrid star) or a completely hadronic phase (neutron star). \par 

Notably, all the energy conditions, as shown in Figs.~\ref{fig18} and \ref{fig19} for the parametric choice of various model parameters, are well satisfied in the model, indicating the validity of the model. Figs.~\ref{fig20}-\ref{fig21} represent the equilibrium under the combined action of gravity force ($F_g$), hydrostatic force ($F_h$) and force due to modified gravity ($F_\zeta$). In Figs.~\ref{fig22} and \ref{fig23} the radial variation of the adiabatic index ($\Gamma$) is plotted for different suitable choices of model parameters, and it is evident that for all the choices, the condition $\Gamma>\frac{4}{3}$ is satisfied. The stability of the present model under small radial disturbances has been examined by plotting the absolute Lagrangian change in radial pressure at the stellar boundary against the eigenfrequencies ($\omega_0^2$) of various oscillation modes as shown in Fig.~\ref{fig24}. The presence of real frequencies ($\omega_0^2>0$) in the spectrum confirms the model's stability. Additionally, the external gravitational effects have been investigated by calculating the tidal Love number ($k_2$) and tidal deformability ($\Lambda$) for different compact stars with the value of the predicted radius from the model as mentioned in Table~\ref{tab3}. All the results are listed in Table~\ref{tab4}. The variation of tidal deformability ($\Lambda$) with coupling parameter ($\zeta$) and with baryon number density ($n$) within stable window are shown in Figs.~\ref{fig25} and \ref{fig26} for the chosen compact object $4U~1820-30$ with mass $1.58~M_{\odot}$, respectively. Notably, for the parameter space used here, the value of tidal deformability ($\Lambda$) lies below the value $800$ \cite{BPA,ABA}. Again, as $m_s$ increases, $\Lambda$ decreases for fixed $\zeta$ which is evident from Fig.~\ref{fig25}. Further, when $n$ increases, $\Lambda$ also increases as shown in Fig.~\ref{fig26}. Since increasing $n$ lowers the value of $E_B$ and hence the energy density, which in turn also lowers the pressure, i.e., the EoS becomes softer. Therefore, with the increase of $n$, the compressibility increases and system becomes prone to higher tidal deformability. Also for a given $\zeta$ and $n$, $\Lambda$ decreases with the increase in $m_s$. This may be attributed as larger strange quark mass tends to steeper the EoS, i.e., high pressure is required for a given energy density. So a steeper EoS results in more compact star, thus reducing tidal deformability. Therefore, for stability, a higher value of $m_s$ is desirable. Hence, the present model is found to characterise the mass-radius relationship and internal structure of a diverse range of SQS candidates. Inclusion of baryon number density $(n)$ dependent bag model in $f(R,T)$ gravity provides some interesting results. Density dependent bag softens the EoS at higher densities, often reducing stellar radius and tuning mass-radius relation closer to observational data. Geometry-matter coupling $(\zeta)$ offers flexibility. Positive $\zeta$ tends to shrink stars and reduce maximum mass, negative value of $\zeta$ could increase compactness depending on model details. Such frameworks enhance modelling of strange stars, circumvent potential singularities and allow clearer tuning to match compact stars like $PSR~J0030+451$, $VELA~X-1$ etc. Additionally, within the framework of $f(R,T)$ gravity, the existence of a stable configuration for compact stars may be possible in this model.

\section*{Acknowledgments}
RR is thankful to the Department of Physics, Coochbehar Panchanan Barma University, for providing necessary help to carry out the research work. DB is thankful to the Department of Science and Technology (DST), Govt. of India, for providing the fellowship vide no: DST/INSPIRE/Fellowship/2021/IF210761. PKC gratefully acknowledges support from IUCAA, Pune, India, under the Visiting Associateship Programme.








\end{document}